\documentclass[12pt,preprint,numberedappendix,twocolappendix]{emulateapj}

\newcommand\bibinc{n}		

\usepackage{subeqnarray}
\usepackage{amsmath}
\usepackage{hyperref}
\usepackage{ulem}
\usepackage{stmaryrd}

\hypersetup{colorlinks=true,linkcolor=red,citecolor=blue,urlcolor=red}

\usepackage{natbib}
\def\tt{  }
\def\ttt{}
\def\bbb{}

\newcommand{\trad}{\tau_{\rm{rad}}}

\setcitestyle{citesep={,}}
\begin{document}

\slugcomment{Submitted to AAS Journals}

\shorttitle{Atmospheric circulation of close-in gas giants}
\shortauthors{Tan \& Showman}

\title{Atmospheric circulation of tidally locked gas giants with increasing rotation and implications for white-dwarf-brown-dwarf systems}
\author{Xianyu Tan$^{1,2}$ and Adam P. Showman$^{2,3}$}
\affil{$^1$Atmospheric Oceanic  and Planetary Physics, Department of Physics, University of Oxford, OX1 3PU, United Kingdom \\
\url{xianyu.tan@physics.ox.ac.uk}\\
$^2$Lunar and Planetary Laboratory, University of Arizona, 1629 University Boulevard, Tucson, AZ 85721 \\
$^3$Department of Atmospheric and Oceanic Sciences, Peking University, Beijing, People’s Republic of China \\
}
\begin{abstract}
 Tidally locked gas giants, which exhibit a novel regime of day-night thermal forcing and extreme stellar irradiation, are typically in several-day orbits, implying a modest role for rotation in the atmospheric circulation. Nevertheless, there exist a class of gas-giant, highly irradiated objects---brown dwarfs orbiting white dwarfs in extremely tight orbits---whose orbital and hence rotation periods are as short as 1-2 hours. Phase curves and other observations have already been obtained for this  class of objects, raising fundamental questions about the role of increasing planetary rotation rate in controlling the circulation. So far, most modeling studies have investigated rotation periods exceeding a day, as appropriate for typical hot Jupiters. In this work we investigate atmospheric circulation of tidally locked atmospheres with decreasing rotation periods (increasing rotation rate) down to 2.5 hours. With decreasing rotation period, we show that the width of the equatorial eastward jet decreases, consistent with the narrowing of { the equatorial waveguide} due to a decrease of the equatorial deformation radius. The eastward-shifted equatorial hot spot offset decreases accordingly, and the off-equatorial westward-shifted hot areas become increasingly distinctive. At high latitudes, winds become weaker and { more} rotationally dominated. The day-night temperature contrast becomes larger due to the stronger influence of rotation. Our simulated atmospheres exhibit variability, presumably caused by  { instabilities and wave interactions}. Unlike typical hot Jupiter models, thermal phase curves of rapidly rotating  models show a near alignment of peak flux to secondary eclipse.  This result helps to explain why, unlike hot Jupiters, {\ttt brown dwarfs closely orbiting white dwarfs tend to exhibit IR flux peaks nearly aligned with secondary eclipse.} Our results have important implications for { understanding} fast-rotating, tidally locked atmospheres. 
\end{abstract}

\section{introduction}

\subsection{Observational motivation}
Since the first transiting observation of HD 209458 b \citep{charbonneau2000},  gas giant planets with small semi-major axes (a.k.a. the hot Jupiters) remain the best characterized group of exoplanets. Due to their close-in  distances, they experience extreme stellar irradiation and, because they are expected to be tidally locked, exhibit a novel regime of permanent day-night thermal forcing.  Thermal phase curves and secondary eclipse  measurements have been obtained for a variety of hot Jupiters, providing constraints on their three-dimensional (3D) thermal structure and circulation pattern \citep[][]{heng2015}. The flood of observations have stimulated a growing body of modeling studies on the 3D atmospheric circulation of  tidally-locked gas giants (e.g., \citealp{showman2002, cooper2005, DobbsDixon2008, showman2009, menou2010, heng2011b, perna2012, rauscher2013, mayne2014, mendonca2016, roman2019, steinrueck2019}). Yet most observations and theoretical models to date have focused on planets that are typically in several-day orbits, implying low rotation rate and a modest role for rotation in the atmospheric circulation. 

Nevertheless, there exist a class of gas-giant, highly irradiated objects---brown dwarfs (BDs) orbiting white dwarfs (WDs) in extremely tight orbits---whose orbital and hence rotation periods can be as short as 1 to 2 hours. These extremely  close-in gas giants are expected to be synchronously rotating as the tidal spin-down time is short compared to system ages (e.g., \citealp{guillot1996}). Thus their orbital period provides a good estimate of the rotation period. These systems are often  survivors of binary  evolution---when sun-like stars become old and expand, engulfing their close-by companions, the orbits of their smaller companions shrink due to friction. The outer layer of the evolved star eventually puff off, leaving a remnant white dwarf with a closely orbiting low-mass (i.e., brown dwarf) companion \citep{hellier2001, percy2007}. Many secondary brown dwarfs  of such systems are donors, i.e., materials from their atmospheres are accreted onto the white dwarfs, and their observations are often complicated by contributions from the accretion discs as well as distinctly non-spherical shape of the brown dwarfs (e.g., \citealp{santisteban2016}).   Fortunately, some of them are detached systems in which there are no outflows from the secondary brown dwarfs, making them the ideal targets to understand the 3D  atmospheric structure and circulation of gas giants that are extremely fast rotating and highly irradiated.  An observational advantage of  this class of objects is that near-infrared photons emitted  from the companion brown dwarfs are generally well separated in wavelength from those emitted from the host white dwarf (whose flux peak is typically in the UV); typically the brown dwarf emits far greater IR energy than the white dwarf, greatly aiding observational characterization.\footnote{This differs from typical hot Jupiter systems, where the star emits thousands of times more radiation than the hot Jupiter even in IR wavelengths where the hot Jupiter's radiation peaks, making it harder to disentangle the planet's signal from the star's signal in noisy observations.}  Phase curves  and other  observations have been obtained for several  such systems, including  NLTT 5306 (orbital period 102 min,   \citealp{steele2013}),  WD0137-349 (116 min, \citealp{casewell2015, longstaff2017}),  EPIC 21223532 (68 min, \citealp{casewell2018}), WD 1202-024 (71 min, \citealp{rappaport2017}) and SDSS J141126.20+200911.1 (122 min, \citealp{littlefair2014,casewell2018b}). In particular, phase curves of WD0137-349 and SDSS J141126.20+200911.1 were taken with multiple wavelength.     Several candidate detached WD+BD systems should be observable in the near future (see a summary in \citealp{casewell2015}).   These phase curves often exhibit  large day-night temperature differences and  nearly zero phase offsets between the peak flux and phase 0.5 where a secondary eclipse would occur if the system were eclipsing. In some cases the flux peaks slightly after phase 0.5, which is in contrast to most hot-Jupiter near-IR phase curves  that show robust peak flux before the secondary eclipse. This   raises fundamental questions about the role of strong rotation in controlling the global circulation and day-night heat transport in tidally locked atmospheres.

In between the extremely close-in brown dwarfs around white dwarfs and the canonical hot Jupiters which are typically in several-day orbits,  a handful of giant planets and brown dwarfs with orbital periods close to or less than a day around sun-like stars have been discovered or characterized (e.g., WASP-12b --- orbital period 1.1 day, \citealp{swain2013}; WASP-103b --- 0.93 day, \citealp{cartier2016, kreidberg2018}; WASP-18b --- 0.94 day, \citealp{sheppard2017,arcangeli2018}; WASP-19b --- 0.78 day, \citealp{espinoza2018}; NGTS-7Ab --- 0.68 day, \citealp{jackman2019}; TOI 263.01 --- 0.56 day, \citealp{parviainen2019}). There are significant observational motivations to   investigate  atmospheric physics for this emerging class of exoplanets (e.g., \citealp{bell2018,parmentier2018,lothringer2018, kitzmann2018, komacek2018rnaas, tan2019b}).  In the near future, global atmospheric properties and circulation of these short-period planetary companions will be better constrained via phase curves and secondary eclipse measurements, and therefore a better  understanding of the stronger role of rotation in atmospheric dynamics is needed.

\subsection{Theoretical motivation}
\label{ch.expectation}

\begin{figure}      
\epsscale{1.1}  
\plotone{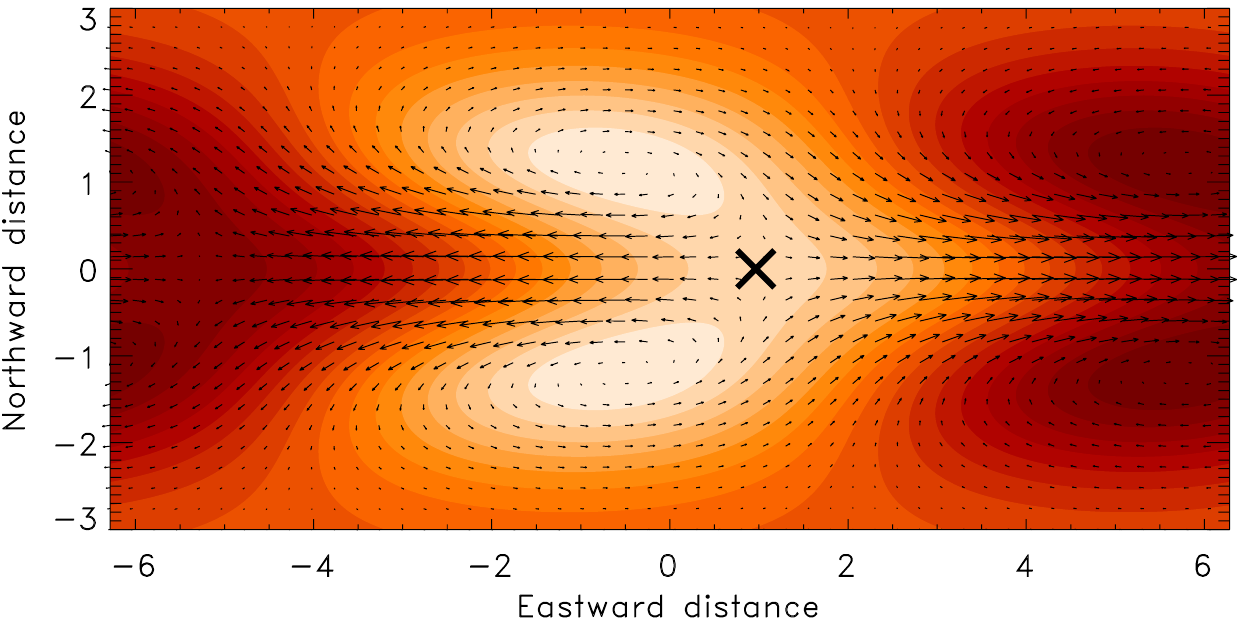}
\caption{Height fields (with arrows representing wind vectors overplotted) of the wave solutions triggered by day-night forcing for a shallow-water system from \cite{showman2011}. { The solution adopts finite $\tau_{\rm{drag}}$ and $\tau_{\rm{rad}}$ which are appropriate for typical hot Jupiters. At the high day-night forcing amplitudes relevant to hot Jupiters, nonlinear effects are important and in that case advection can play the qualitative role of drag, leading to a similar structure even when there is no drag in the active layer. Note that for typical hot Jupiters, the structure extends all the way from the equator to high latitudes. At faster rotation rate, a similar structure occurs, but becomes confined closer to the equator due to the decrease in equatorial deformation radius with increasing rotation rate.}}
\label{closebd.sp11wave}
\end{figure} 

The dominant dynamical { response to day-night radiative forcing} in the atmosphere of a canonical hot Jupiter is the excitation of standing, global-scale equatorial Rossby and Kelvin waves. The superposition of the equatorial Rossby and Kelvin waves (the so-called Matsuno-Gill pattern, \citealp{matsuno1966, gill1980}) causes phase tilts in the eddy velocities that pump eastward momentum from high latitudes to the equator, thereby inducing a fast eastward equatorial jet, or superrotation. { Combined with advection due to the equatorial superrotation}, the eastward propagation of the Kelvin waves also help displace the hot spot eastward of the substellar point.  Analytical wave solutions were given by \cite{showman2011} in a linearized shallow-water system with finite radiative damping and frictional drag. Subsequently, \cite{tsai2014} investigated solutions in a continuously stratified atmosphere with a uniform zonal wind in the regime of equal radiative and drag timescale; \cite{heng2014} extended the shallow-water solutions to include magnetic field and spherical geometry; { and} \cite{hammond2018} investigated effects of general profiles of zonal-mean zonal wind on the wave-mean-flow interactions using the shallow water system; \cite{debras2019} explored the time-dependent wave solutions under day-night forcing. An example of the steady wave solution  with  moderate non-dimensionalized radiative and drag timescale which are  appropriate for typical hot Jupiters  is shown in the upper panel of Figure \ref{closebd.sp11wave}.  In this regime, the northwest-southeast (northeast-southwest) eddy tilt in the northern (southern) hemisphere pumps eastward angular momentum from the off-equatorial region to the equatorial region.  A general understanding of the emergence of this eddy tilt is the three-way horizontal force balance between pressure-gradient, Coriolis and frictional drag force. Even when the explicit frictional drag is weak as may be the case in many hot Jupiters, similar three-way force balance can still be achieved by nonlinear advection terms, allowing emergence of such a  eddy tilt and driving the equatorial superrotation  ({ \citealt{showman2011} and \citealt{showman2013}; for a brief review see \citealp{showman2013b}}). 
   
This Matsuno-gill pattern is typical in slow rotators in which the Rossby number $Ro\sim1$ and the equatorial Rossby deformation radius $L_D$ is comparable to planetary radius.  The deformation radius is defined as $L_D = \sqrt{c/\beta}$, where $c$ is gravity wave speed and $\beta=df/dy$ is the meridional gradient of the Coriolis parameter $f$, and $y$ is distance, { increasing} northward. The Coriolis parameter is $f=2\Omega\sin\phi$, where $\phi$ is latitude and $\Omega$ is { the planetary rotation rate}, and therefore $\beta=2\Omega\cos\phi/a$, where $a$ is { planetary} radius, implying $\beta=2\Omega/a$ at the equator. A good estimate of the horizontal phase speed of long-vertical-wavelength gravity waves is $c\sim2NH$ where $N$ is Brunt-Vaisala frequency and $H$ is scale height. One can then express the equatorial deformation radius as $L_D=\sqrt{NHa/\Omega}$. The theory of \citet{showman2011} predicts that the meridional half-width of the equatorial jet on synchronously rotating planets is approximately the equatorial deformation radius.

As the rotation period decreases (rotation rate increases), the deformation radius decreases relative to the planetary radius, and the meridional extent of the associated standing waves  are expected to be smaller and smaller, confining the wave structure closer to the equator. At high latitudes{---beyond the equatorial waveguide---}the force balance is expected to be primarily between the pressure-gradient and the Coriolis force if the frictional drag is weak, which is the so-called geostrophic regime. 


Based on the above framework, we expect to see an emergence of two { distinct} behaviours with decreasing rotation period---standing waves are triggered but confined closer to the equator, and at high latitudes the flow becomes increasingly geostrophic. Equatorial superrotation may still be prevalent but its meridional extent would decrease due to the smaller equatorial deformation radius. These changes in the  circulation pattern will directly influence the thermal phase curves and other observables. In the slow-rotating regime, the eastward-shifted hot spot dominates and results in a peak thermal flux before the secondary eclipse. However in { rapidly} rotating cases, when the Matsuno-Gill pattern is confined closer to the equator, the equatorial jet and the eastward shifted hot spot occupy a smaller fraction of the planet's area. {\ttt Meanwhile, the  westward-shifted hot regions associated with the off-equatorial Rossby gyres (see the off-equatorial height maximum in Figure \ref{closebd.sp11wave}) could contribute significantly to thermal phase curve, therefore affecting the phase of the peak flux.} This effect has been   shown by \cite{penn2017} using shallow water models.   Flows tend to be  geostrophic in the rapid rotating regime which helps to sustain large horizontal temperature differences against heat transport by wave adjustment, and as a result the fractional day-night temperature difference is expected to increase with decreasing rotation period \citep{perezbecker2013, komacek2016}. 

Previous work has explored the influence of  rotation rate on the atmospheric circulation of hot Jupiters.  \cite{showman2008} varied rotation rate under the assumption of synchronous rotation, while \cite{showman2009} and  \cite{rauscher2014} explored  rotation rate varying simultaneously with allowing the heating pattern to migrate non-synchronously, all within a factor of two or so variations in rotation rate. \cite{showman2015} widened the range of rotation rate variations  including non-synchronous rotation. \cite{komacek2017} varied rotation rate and stellar  insolation in a self-consistent manner. \cite{kataria2013} performed simulations for quasi-fast-rotators in the context of eccentric planets rotating pseudo-synchronously. \cite{penn2017} explored effects of different rotation rate and nonsynchronization on the thermal phase curve offset using shallow-water models. {\ttt \cite{flowers2019} systematically compared observed high-resolution transmission spectrum of HD 189733b to models with a wide range of rotation period.  } In general, these studies tend to support the picture that faster rotation leads to a narrower equatorial jet. But in most of these studies, the complicating effects of { non-synchronous}  rotation, varying stellar insolation and eccentric orbits complicate an understanding of the effect of varying rotation by itself.  More importantly, none of these studies reach the ultra-fast-rotation regime relevant to white-dwarf-brown-dwarf binaries, as we want  to do here.  

Time-dependent  processes, including large-scale instability and free-propagating waves, have been given little attention in existing theoretical work for hot Jupiters, partly because, in both observations and numerical simulations of hot Jupiters, the dominant atmospheric features are almost time invariant. A well-known source of time-variability in  rapidly rotating atmospheres is baroclinic instability, which dominates the large-scale mid-latitude dynamics in Earth's troposphere (for reviews see, e.g., \citealp{showman2013, vallis2017}), and may contribute to driving the multiple zonal jets in Jovian and Saturnian atmospheres (see a recent review by \citealp{showman2019review}).  Conditions { on} the close-in gas giants differ from low-temperature terrestrial  atmospheres in several ways, including a much shorter radiative timescale near the photospheres, the possibly strongly stratified photospheres,  the lack of a surface temperature gradient, and finally the day-night thermal forcing instead of the equator-to-pole forcing more typical on planets like Earth, Mars, and the Solar system's giant planets. At present it is therefore unclear whether (and to what extent) baroclinic instability { plays} an important role in maintaining the atmospheric circulation of close-in gas giants. Other types of large-scale instability---for example barotropic instability, { which} is relevant in Jovian and Saturnian atmospheres \citep{ingersoll1982}---are also worth examining in the context of day-night-forced, rapidly rotating atmospheres. \\

To gain a comprehensive understanding of the  circulation on rapidly rotating, tidally locked gas-giant planets, it is best to seek solutions using well-controlled numerical experiments.  In this study, utilizing a general circulation model (GCM) with idealized day-night thermal forcing, we investigate effects of increasingly strong rotation on the atmospheric circulation of tidally locked atmospheres.  This will  help to provide a foundation for understanding current and future observations of ultra-close-in, rapidly rotating brown dwarfs and exoplanets. To isolate the dynamical effects of rotation in a clean, systematic manner,  we intentionally simplify our model by excluding other potentially important effects on the circulation, such as feedbacks involving radiatively active clouds (e.g.,   \citealp{lee2016, lines2018, roman2019}), magnetic drag (e.g., \citealp{perna2010, batygin2013, rauscher2013, rogers2014komacek, rogers2017}), sophisticated interplays between radiation and chemistry at high temperature \citep{parmentier2018, lothringer2018} and heat transport by hydrogen dissociation and recombination \citep{tan2019b, komacek2018rnaas, bell2018}. In addition, vigorous atmospheric circulation of field brown dwarfs has been inferred by numerous observations (see a recent review by \citealp{biller2017}).  The circulation on these field brown dwarfs is solely driven by the internal  heat flux, which presumably does not have day-night or equator-pole variation \citep{showman&kaspi2013}.  Yet these objects likewise exhibit fast rotation, high temperature and high surface gravity.    The use of  a simplified GCM here offers a clean comparison to investigations of isolated brown dwarfs using idealized global models \citep{zhang&showman2014, tan2017, showman2019} whose atmospheric circulation  is forced in a very different manner. 


This paper is organized as follows. We introduce the numerical model in Section \ref{ch.model}. In Section \ref{ch.results}, we first describe general results for models with short radiative timescale in Section \ref{ch.result1} and for models with long radiative timescale in Section \ref{ch.result2}. We explore effects of drag on rapid rotators in Section \ref{ch.result4} and present synthetic phase curves of models in the grid in  Section \ref{ch.result5}.   Finally, we discuss implications of our results in Section \ref{ch.discuss} { and highlight key
conclusions in Section~\ref{ch.conclude}.}

\section{Model}
\label{ch.model}

We solve the standard global hydrostatic primitive equations in pressure coordinates, which include the horizontal momentum, hydrostatic equilibrium, continuity and thermodynamic energy equations. The ideal gas law is assumed for the equation of state of the atmosphere.    Our { physical} model is almost identical to those used in \cite{liu2013} and \cite{komacek2016}, where  we refer readers for more details. Radiative heating and cooling are represented { using} a Newtonian scheme in which the temperature field is relaxed towards a prescribed equilibrium temperature profile over a characteristic radiative timescale $\trad(p)$. The radiative timescale $\trad$ is a function of pressure only.  The radiative-equilibrium temperature {\tt structure} $T_{\rm{eq}}(\lambda, \phi, p)$, { which is hot on the dayside and cold on the nightside, is taken to be the following} function of longitude $\lambda$, latitude $\phi$ and pressure $p$: 
\begin{equation}
 T_{\rm{eq}}(\lambda, \phi, p) = \left\{
\begin{array}{lr}
T_{\rm{night, eq}}(p) + \Delta T_{\rm{eq}}(p) \cos \lambda \cos \phi \quad \rm{dayside} \\
T_{\rm{night, eq}}(p) \quad \rm{nightside.}
\end{array}
\right.
\end{equation}
Here $T_{\rm{night, eq}}(p)$ is the  equilibrium profile on the nightside and $T_{\rm{night, eq}}(p)+\Delta T_{\rm{eq}(p)}$  is that at the substellar point. The nightside profile is acquired by subtracting our chosen $\Delta T_{\rm{eq}(p)}/2$ from a temperature profile of HD 209458b from \cite{iro2005}. { The day-night difference in radiative-equilibrium temperature,} $\Delta T_{\rm{eq}}(p)$, is set to be a constant $\Delta T_{\rm{eq, top}}$ at pressure lower than $p_{\rm{eq, top}}$, zero at pressure larger than $p_{\rm{bot}}$ and varying linearly with log-pressure in between.  We take $p_{\rm{eq, top}}=10^{-3}$ bar, $p_{\rm{bot}}=10$ bars and $\Delta T_{\rm{eq, top}}=1000$ K for most of the models. The radiative timescale is generally short in the upper atmosphere and long at depth (e.g, \citealp{iro2005, showman2008}). To qualitatively capture this behaviour, $\trad$ is set to a small constant $\tau_{\rm{rad, top}}$ at pressures less than $p_{\rm{rad, top}}$ and a large constant $\tau_{\rm{rad, bot}}$ at pressures greater than $p_{\rm{bot}}$, then varies linearly in logarithmic space in between.  We set $p_{\rm{rad, top}}=10^{-2}$ bar and $p_{\rm{rad, bot}}=10$~bar.  This forcing setup is nearly identical to that in \cite{liu2013} and \cite{komacek2016}. 
 { Because brown dwarfs in BD+WD systems experience a wide range of radiative time constants, and because} there is a strong theoretical motivation for investigating {\tt how the} atmospheric circulation { responds to} varying radiative timescale \citep{showman2011,perezbecker2013, komacek2016},
{ we systematically explore values of $\tau_{\rm{rad, top}}$ ranging from $10^4$ to $10^7\rm\,s$.  Rotation period is likewise varied over a wide range.} 
 
The frictional drag is represented by a linear damping of horizontal velocities, which represents missing physics, for instance, turbulent mixing or the Lorentz force on the velocity (see \citealp{komacek2016} for a detailed discussion). Kinetic energy dissipated by the drag is returned to the system as a heating term in the thermodynamic energy equation.  As in \cite{komacek2016}, there are two components in the drag. The first one is a basal drag applied only near the bottom of the model domain, which crudely parameterizes interactions between the weather layer and a relatively quiescent planetary interior. The basal drag is zero (meaning no drag) at pressure less than $p_{\rm{drag, top}}$, and the drag coefficient linearly increases with increasing pressure from zero at pressure $p_{\rm{drag, top}}$ to a maximum value $1/\tau_{\rm{drag,bot}}$ at the bottom of the domain ($\sim$200~bars). At the bottom of the computational domain the basal drag strength is characterized by a drag timescale $\tau_{\rm{drag, bot}}$. We set $p_{\rm{drag, top}}=10$ bars and $\tau_{\rm{drag, bot}}=10^6$ s for all models. The basal drag does not directly influence the dynamics at pressure less than 10 bars. The second component, which we only include in some models, is a spatially independent drag{---applied everywhere in the domain---}that is characterized by a constant drag timescale $\tau_{\rm{drag}}$.  It is intended primarily as a very crude representation for Lorentz forces that might be associated with partial ionization on particularly hot planets (e.g., see \citealp{komacek2016}, \citealp{rauscher2013} and \citealp{perna2010} for discussion).  Note that most of our models lack the second component; however, we will present some models that include it. The total frictional drag at any grid point takes the maximum of the two drag components.
 
We adopted planetary parameters relevant for gas giants, including specific heat $c_p = 1.3 \times 10^4 ~\rm{J~kg^{-1}~K^{-1}}$, specific gas constant $R =  3714 ~\rm{J~kg^{-1}~K^{-1}}$ and a planetary radius  $7\times 10^7$ m (similar to that of Jupiter).  Because the major motivation of this study  comes from observations of brown dwarfs around white dwarfs, we adopted surface gravity $g=1000 ~\rm{m~s^{-2}}$  relevant for brown dwarfs. However, as discussed in \cite{showman2019}, the entire system is independent of the value of gravity for our particular model formulation (i.e., the equations are solved in pressure coordinates with a prescribed Newtonian cooling scheme that is also specified to be a function of pressure rather than height). Thus, the qualitative implications from our numerical results should be applicable to both hot Jupiters and brown dwarfs. We systematically vary the rotation period from 80 to 2.5 hours and $\tau_{\rm{rad, top}}=10^4$  to $10^7$ s, and assume a drag-free  atmosphere for the main suite of simulations (in other word, our main suite of simulations has only the basal drag scheme at pressures greater than 10 bars; the main atmosphere at $p<10$ bars lacks any frictional drag). In some rapidly rotating experiments, we include finite drag timescale $\tau_{\rm{drag}}$ throughout the atmosphere  to explore dynamics with increasing drag strength. 

We use the MITgcm to solve the global hydrostatic primitive equations using the cubed-sphere grid \citep{adcroft2004}. A standard fourth-order Shapiro filter is applied to the horizontal momentum and thermodynamics equations to maintain numerical stability, and it has minimal effect on the large-scale structures. The pressure domain is between 200 bars and $2\times10^{-4}$ bar and is evenly divided into 40 levels  in log pressure. Because the dynamical length scales (e.g., deformation radius) are small at fast rotation, properly resolving the global-scale dynamics at fast rotation requires greater horizontal resolution than at slow rotation.  As is standard in the field, we adopt a resolution for our slowly rotating, 80-hour and 40-hour models of C32 (meaning  $32 \times 32$ points on each cubed-sphere face, equivalent to { a global resolution of} $128\times64$ in longitude and latitude)  but increase the resolution up to C192 for models at a 2.5-hour rotation period.  The latter implies that we use $192 \times 192$ points on each cubed-sphere face, equivalent to a global resolution of approximately $768 \times 384$ in longitude and latitude, or $0.47^\circ$ of longitude or latitude  per grid cell.  These latter  models are among the highest-resolution models of tidally locked giant planets that have ever been performed.\footnote{\cite{liu2013} showed, for a canonical hot Jupiter with a rotation period of 3.5 days and a forcing setup similar to that here, that the statistically equilibrated final state obtained using a horizontal  resolution of C32 is almost identical to that obtained at a much higher resolution of C128, implying that C32 is adequate for understanding many aspects of  the system under the slowly rotating conditions of canonical hot Jupiters.   {\cite{menou2019} also showed that for canonical hot Jupiters, modest resolution is sufficient to resolve the main dynamical structure.} However, C32 would be insufficient for the faster-rotation models considered in this paper.}

All models are integrated to a statistical steady state. The initial state is at rest with a globally uniform temperature profile set to the \cite{iro2005} profile. Note that equilibrated final state is insensitive to initial conditions \citep{liu2013}.

\section{Results}
\label{ch.results}
We performed simulations in a grid with varying rotation period 80, 40, 20, 10, 5 and 2.5 hours and varying radiative timescale at the top $\tau_{\rm{rad, top}}=10^4$, $10^5$, $10^6$ and $10^7$ s. Simulations in this grid are drag-free at pressure lower than 10 bars, and a basal drag at pressure higher than 10 bars is included. Although $\tau_{\rm{rad, top}}$ of some models are higher than that physically motivated especially for rapid rotators, understanding dynamics in the high-$\tau_{\rm{rad}}$ regime is fruitful to obtain a full dynamical picture,  as has been by previous studies of canonical hot Jupiters \citep{perezbecker2013, komacek2016,komacek2017}. Our grid connects to the that in \citet{komacek2016} in which they explored the circulation with different drag strength and radiative timescale but with a fixed rotation period appropriate for canonical hot Jupiters.  

\begin{figure*}      
\epsscale{1.15}      
\plotone{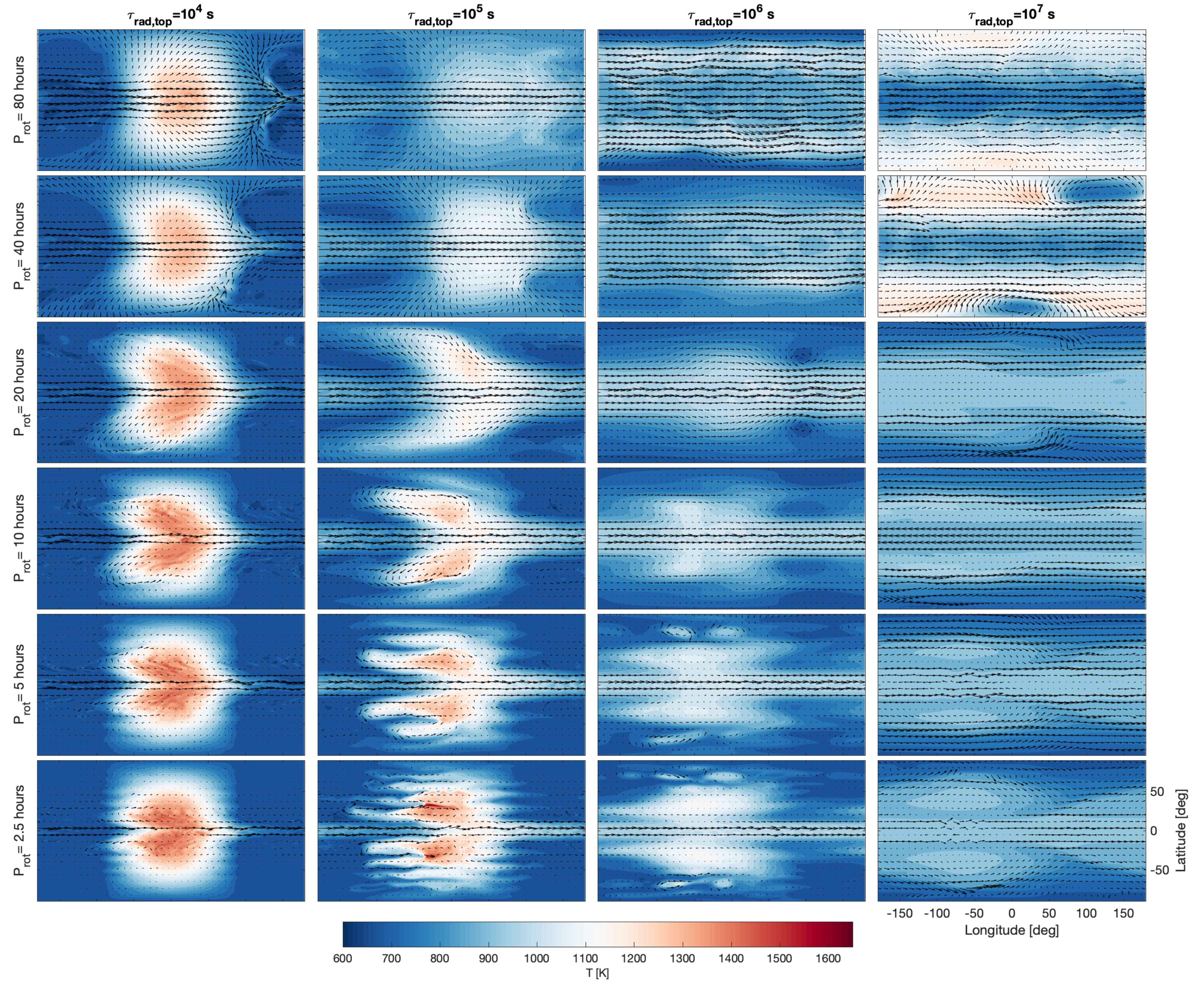}
\caption{Snapshots {showing the global temperature structure (colorscale, in K) and overplotted wind vectors (arrows)} at 7~mbar pressure { for a full grid of} numerical experiments with different rotation period (along the vertical direction) and different radiative timescale $\tau_{\rm{rad,top}}$ (along the horizontal direction). { In total, 24 distinct numerical simulations are shown.  In each case, the substellar point lies at a longitude and latitude of $(0^\circ,0^\circ)$, the dayside corresponds to the center portion of each panel, and the planetary terminators are at longitudes $\pm90^\circ$. The longitude and latitude scales are plotted on the panel in the lower right corner, and are the same for all other panels shown.  All models are drag-free at pressures less than 10~bar.} These  snapshots are taken after the models reach statistical equilibrium.}
\label{closebd.temp30}
\end{figure*}

\begin{figure*}      
\epsscale{1.15}      
\plotone{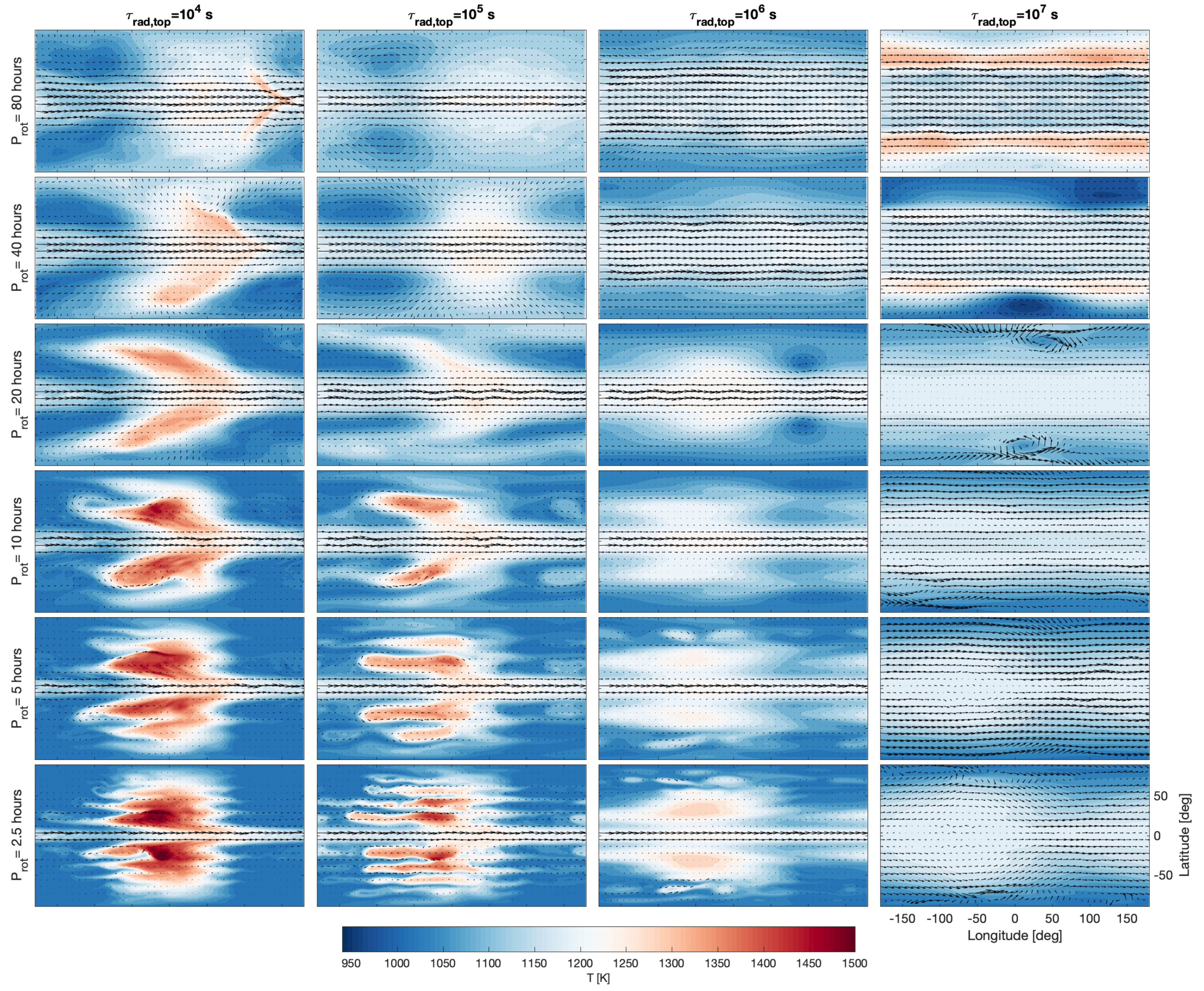}
\caption{Same as Figure \ref{closebd.temp30} but temperature {and winds} at a pressure of 80~mbar.}
\label{closebd.temp23}
\end{figure*}

\begin{figure*}      
\epsscale{1.15}      
\plotone{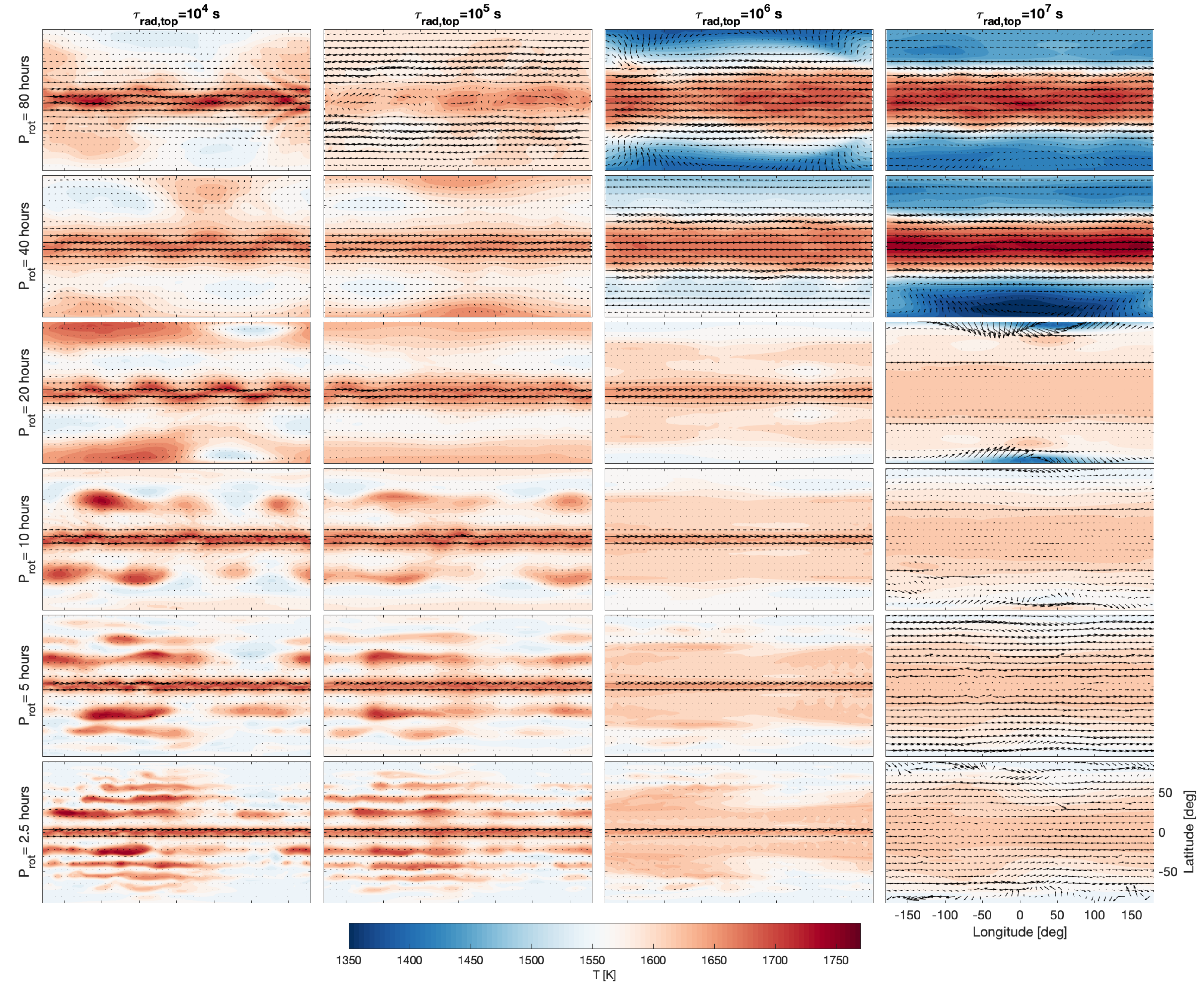}
\caption{Same as Figure \ref{closebd.temp30} but temperature { and winds} at a pressure of 1.3~bar.}
\label{closebd.temp15}
\end{figure*}

\begin{figure*}      
\epsscale{1.15}      
\plotone{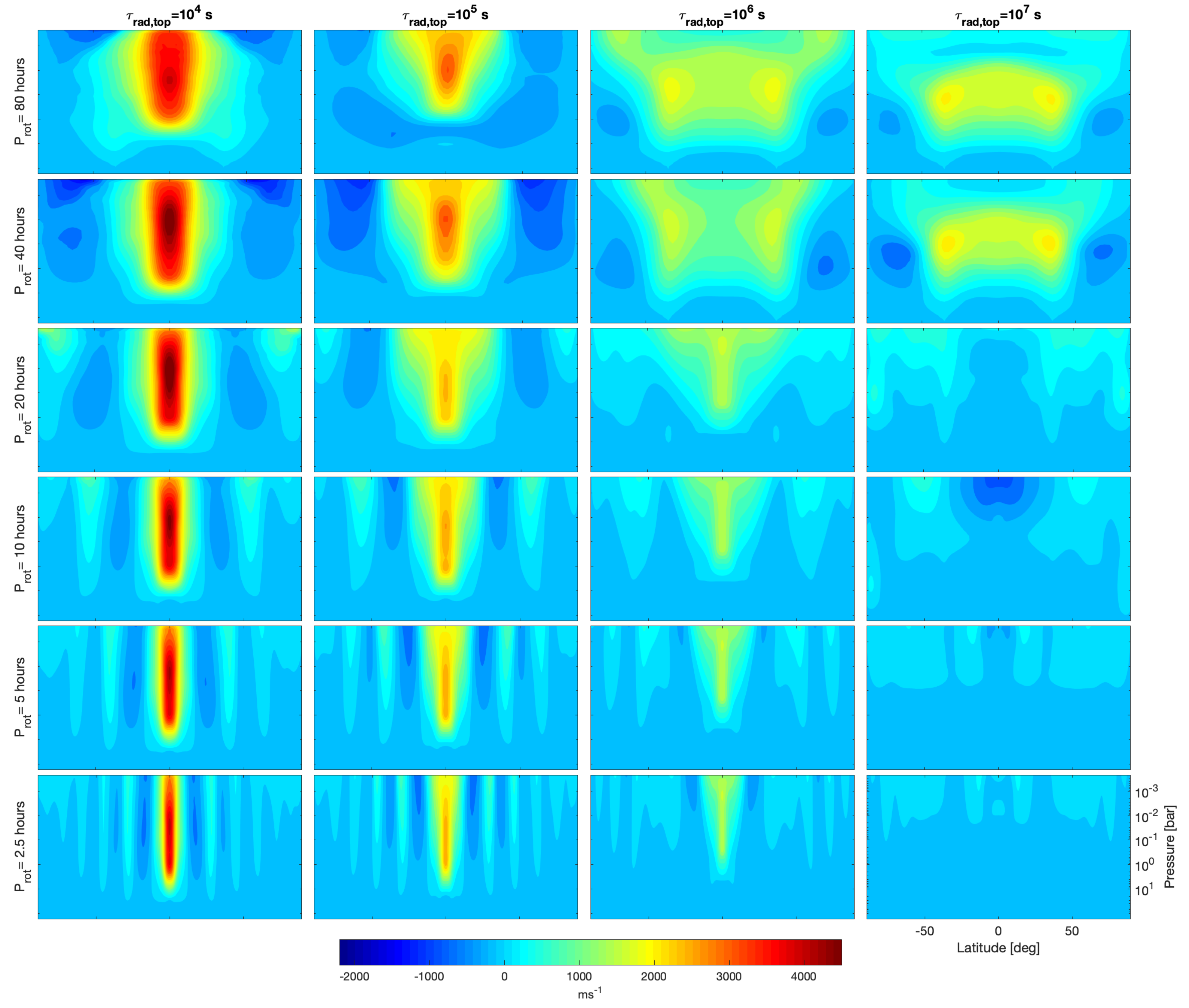}
\caption{Time-averaged zonal-mean zonal wind as a function of latitude and pressure of models over the grid with different rotation period in the vertical direction and different radiative timescale $\tau_{\rm{rad,top}}$ in the horizontal direction. All models are drag-free { at pressures less than} 10~bars.}
\label{closebd.uzonalav}
\end{figure*}

Snapshots of horizontal temperature structure with {superposed} wind vectors for models in the grid are shown in Figure \ref{closebd.temp30} at a pressure 7 mbar, Figure \ref{closebd.temp23} at 80 mbar  and Figure  \ref{closebd.temp15} at 1.3 bar. The time-averaged zonal-mean zonal wind structures are shown in Figure \ref{closebd.uzonalav}. In these plots, the radiative timescale increases from left to right and the rotation period decreases from the top to the bottom.  We first describe the general  outcomes of this grid. Circulation in the canonical hot Jupiter regime (those with rotation period 80 and 40 hours) has been extensively discussed in \cite{komacek2016}, and we focus more on dynamics of the rapidly rotating, short-radiative-timescale models as they are in a novel dynamical regime and relevant for light-curve observations of the WD+BD systems. Some simulations with spatially independent drag will also be discussed. Finally we present synthetic phase curves for models with short radiative timescale.

\subsection{Circulation with short radiative timescale $\tau_{\rm rad,top}=10^4$ and $10^5\rm\,s$}
\label{ch.result1}

{ In the observable atmosphere (pressures less than a few hundred mbar), models with short radiative time constant exhibit a global-scale, large-amplitude day-night temperature difference across a wide range of rotation period. In general,}
the characteristic day-night temperature difference  increases with decreasing rotation period,  { a trend clearly evident in the left two columns of Figures \ref{closebd.temp30} and \ref{closebd.temp23}.  At $\tau_{\rm{rad,top}}=10^4$~s, the radiative timescale is so short at low pressure that the dynamics are inefficient at redistributing heat from day-to-night, and the temperature structure is not far from radiative equilibrium at all rotation rates explored.  Thus, at 7~mbar, the day-night temperature difference increases only mildly with rotation rate (left column of Figure~\ref{closebd.temp30}).  
Interestingly, though, the trend is much more drastic at 80~mbar (left column of Figure~\ref{closebd.temp23}).  This occurs because the dynamics efficiently redistributes the heat from day-to-night when the rotation is slow, but is suppressed from doing so when rotation is fast---leading to very strong dependence of day-night temperature difference on rotation rate.}  For models with $\tau_{\rm{rad,top}}=10^5$ s, the systematic increase of day-night temperature difference with decreasing rotation period { is strong throughout the observable atmosphere, and is prominent at both 7~mbar and 80~mbar in Figures \ref{closebd.temp30}--\ref{closebd.temp23}.}  { In this case, the longer radiative timescale permits significant dynamical readjustment of the temperature structure when rotation is slow, but this dynamical readjustment mechanism becomes much less efficient when rotation is fast.}

 The day-night temperature difference becomes visibly weaker or even indistinguishable at { pressures exceeding 1~bar} where the radiative timescale is longer and the day-night forcing is weaker.   The global circulation is instead dominated by a zonally symmetric { configuration in} which eddies of various horizontal length scales are embedded. The temperature structure at 1.3~bar is shown as an example in Figure \ref{closebd.temp15}. The tendency of the deep circulation to become { increasingly zonally banded} is not surprising, since when the radiative timescale becomes longer than relevant dynamical timescales at depth, latitude-dependent dynamical processes such as interactions between Rossby waves and { turbulence could overcome the direct day-night forcing and drive a zonally banded circulation} (e.g., \citealp{showman2013b,showman2015}). 

With decreasing rotation period,  the global circulation transitions from the canonical hot Jupiter regime to rotation dominated, geostrophic regime. At relatively low pressure, the global-scale pattern of slow rotators with 40 and 80 hour rotation period is dominated by a broad equatorial eastward jet and the eastward shifted hot spot. The standing Rossby waves extend all the way to the poles and  are Doppler shifted towards east by the strong equatorial jet. This regime has been extensively explored by previous work (see a review by, e.g.,  \citealp{heng2015}). In the case with 20-hour rotation period, the off-equatorial Rossby waves are more robustly developed than that with 80 and 40 hour rotation period. As the rotation period {\tt decreases}, the equatorial deformation radius becomes smaller and{---for rotation periods less than about 10~hours---}is only a small fraction of the planetary radius.   The Matsuno-Gill wave pattern excited by the day-night forcing is thus confined closer to the equator and the poleward edges of the standing Rossby waves { shift} to lower latitudes as the rotation period decreases. Poleward of the Matsuno-Gill pattern, horizontal winds  are much weaker than those at low latitudes, and wind vectors follow the isotherms { more closely (Figures~\ref{closebd.temp30} and \ref{closebd.temp23})}. The Rossby number at mid-high latitude is much less than 1 over the global scale, corresponding to the geostrophic circulation regime.
{ At pressure levels where slowly rotating models exhibit a systematic eastward offset of the global-scale dayside hot regions, this offset visibly decreases with decreasing rotation period.  This is particularly evident at $\tau_{\rm{rad,top}}=10^5\rm\,s$ in Figure~\ref{closebd.temp30} and at both $\tau_{\rm{rad,top}}=10^4$ and $10^5\rm\,s$ in Figure~\ref{closebd.temp23}.}  

{ Off the equator, Rossby numbers can be as small as a few percent in our rapidly rotating models.  For example, in our most rapidly rotating model with a 2.5-hour period, the rotation rate is $\Omega = 7\times10^{-4}\rm\,s^{-1}$, implying a typical value of the Coriolis parameter in midlatitudes of $f\approx 10^{-3}\rm\,s^{-1}$. Adopting a wind speed of $200\rm\,m\,s^{-1}$ and a typical meridional length scale of $10^7\rm\,m$ yields $Ro = U/fL \approx 0.02$.}

{ Interestingly, outside the equatorial regions, the zonal-mean temperature of rapid rotators oscillates non-monotonically as a function of latitude.  This ``fingering'' phenomenon is prominent on the dayside at 7~mbar and 80~mbar, where zonally aligned tongues of colder and warmer air interleave as a function of latitude (Figures~\ref{closebd.temp30} and \ref{closebd.temp23}).  It is also prominent on both the dayside and nightside at deeper levels of $\sim$1~bar (Figure~\ref{closebd.temp15}).} The meridional extent of these zonal bands decreases, and the number of zonal bands increases, with decreasing rotation period.   The meridional { positioning of the zonal bands is consistent between high and low pressures.}

These zonal temperature bands are accompanied by numerous zonal jets, alternating in sign as a function of latitude (see panels in lower left corner of Figure~\ref{closebd.uzonalav}).  { As can be seen from Figures~\ref{closebd.temp30}--\ref{closebd.temp15}}, the off-equatorial eastward jets { are positioned} where the zonal-mean temperature decreases poleward, while the westward jets {are positioned} where the zonal-mean temperature increases poleward.  { This relationship implies, through thermal-wind balance, that the zonal jets gradually decay with increasing pressure. Indeed, just such behavior can be seen in} Figure \ref{closebd.uzonalav}.

Interestingly, this non-monotonic zonal-mean temperature structure of our rapid rotating models differs drastically from Earth's troposphere and many GCMs for rapidly rotating giant planets (e.g., \citealp{williams2003, lian2008, schneider2009}), all of which are under zonally symmetric, equator-to-pole thermal forcing and exhibit a monotonic decrease of zonal-mean temperature (on isobars) with increasing latitude.  { The non-monotonic temperature structure seen in our models likely results from the presence of day-night thermal forcing.  One possibility is that the numerous off-equatorial eastward and westward zonal jets induce the temperature fingering by advecting cold nightside air onto the dayside in meridionally localized tongues of air (a mechanism that cannot occur when the forcing is zonally symmetric, as it is in many Jupiter models). Still, the zonal advection timescale for this process\footnote{For a typical zonal jet speed for the off-equatorial zonal jets of $U\approx 200 \rm\,m\,s^{-1}$, and requiring the advection to occur over zonal distances of a planetary radius, leads to an advection timescale $a/U \sim 3\times10^5\rm\,s$, where $a$ is the planetary radius.} is several $\times 10^5\rm\,s$, which may be too long---relative to the radiative time constant---to maintain these temperature variations. 
Another possibility is vertical motion associated with meridional circulations occurring on the small meridional scale of the off-equatorial zonal jets; such motions would advect entropy vertically and induce horizontal temperature contrasts.  Indeed, this mechanism would seem to be preferred at pressures exceeding $\sim$1~bar, where day-night temperature differences are minimal (Figure~\ref{closebd.temp15}).}

\begin{figure}      
\epsscale{1.15}      
\plotone{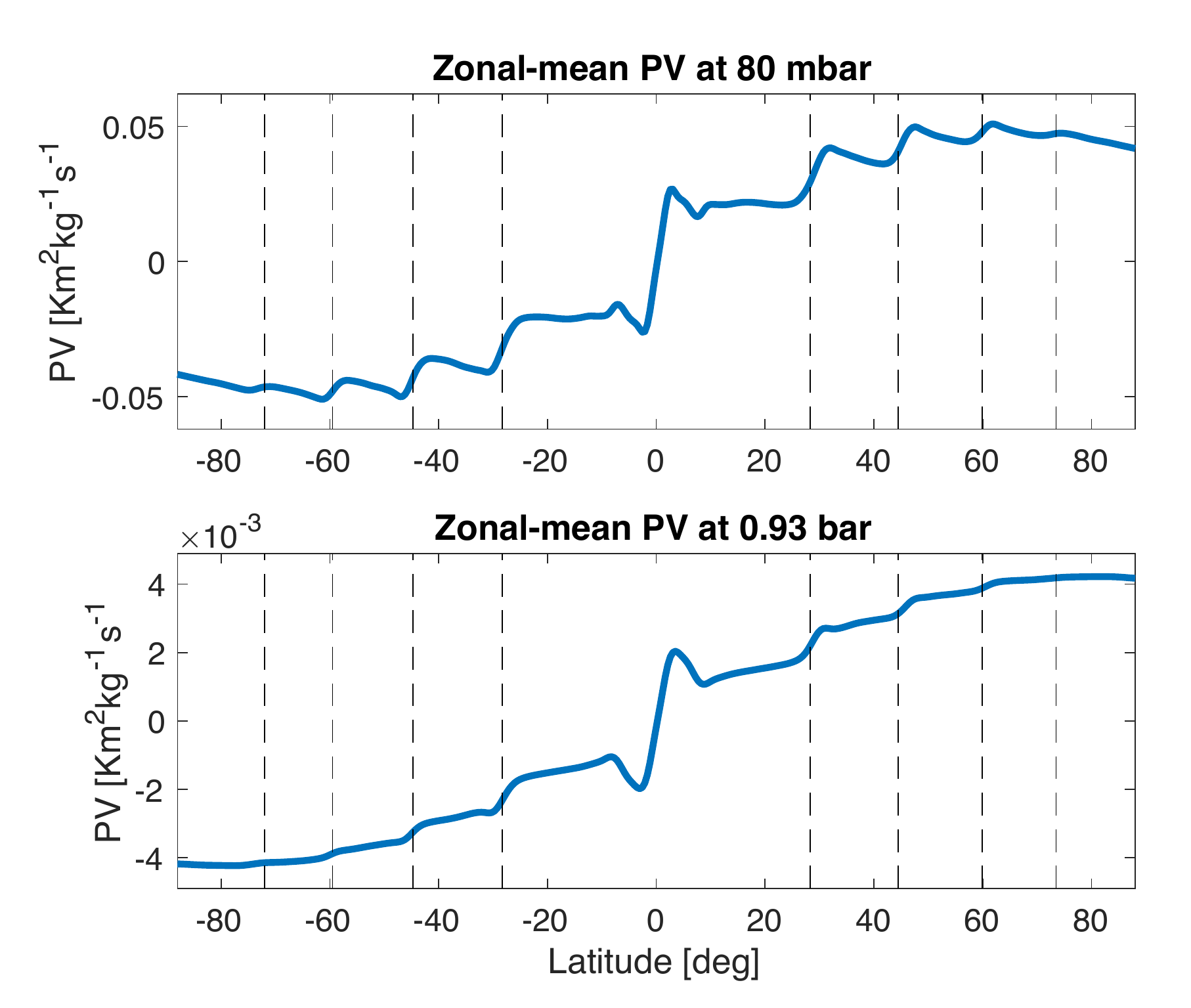}
\caption{{\ttt Zonal- and time-mean potential vorticity (PV) as a function of latitude at two pressure levels (80 mbar in the top panel and 0.93 bar in the bottom panel) for the model with a rotation period of 2.5 hours and $\tau_{\rm rad,top}=10^4$ s. The dashed lines represent latitudes corresponding to  the cores of subtropical eastward jets. }
}
\label{closebd.pv}
\end{figure}

There exist interesting off-equatorial { coherent} structures and small-scale eddies in rapid rotators with rotation period 10 hours and less. Eddies with small horizontal length scale propagate westward at {\tt mid-to-high} latitudes, { an effect which is} especially prominent at 1.3 bar. These eddies have long vertical wavelength and affect { the} temperature structure at pressures lower than 80 mbar, { which is} partly responsible for { the} hot ``tails" west of the dayside meridionally stacked structures there.  The off-equatorial small-scale eddies likely originate from baroclinic instability, { triggering Rossby waves which then propagate westward on the zonal jets}. {\ttt  In the quasi-geostrophic limit,  one of the necessary conditions for baroclinic instability  to occur   is that the meridional gradient of the zonal-mean potential vorticity (PV) changes sign in  the vertical  direction (e.g., \citealp{pierrehumbert1995},  Chapter 6 of \citealp{vallis2006}). The statistically equilibrated zonal-mean configurations of our rapidly rotating models satisfy such a condition. In Figure \ref{closebd.pv}, we show the zonal-mean PV as   a function of latitude at pressure 80 mbar and 0.93 bar for the model with a rotation period of 2.5 hours and $\tau_{\rm rad,top}=10^4$ s. The full hydrostatic  PV  in pressure coordinates shown in Figure \ref{closebd.pv} is written as (e.g., \citealp{vallis2006})
\begin{equation}
    Q=-g\left((f+\zeta)\frac{\partial \theta}{\partial p} - \frac{\partial v}{\partial p}\frac{\partial \theta}{\partial x} + \frac{\partial u}{\partial p}\frac{\partial \theta}{\partial y} \right ),
    \label{eq.pv}
\end{equation}
  where $\zeta=\nabla_p\times \mathbf{v}$ is the vertical component of the relative vorticity, $\mathbf{v}$ is the isobaric velocity vector and $\theta$ is the potential temperature. We have estimated that the Rossby number at mid latitudes is about $Ro\sim 0.02$, and thus the quasi-geostrophic limit holds well at mid latitudes. Poleward of about $\pm 20^{\circ}$, there are regions wherein the magnitude of PV first increases  then decreases poleward (see  regions in between the eastward jet cores marked by the dashed lines). However, at 0.93  bar,  the magnitude of  PV monotonically increases poleward in most latitudes except regions within $\pm10^{\circ}$. This results in a change of sign for $\partial Q/\partial y$ in the vertical direction between these layers, which likely promotes the baroclinic instability at mid latitudes.
}

At low latitudes, transient eddies are also active and { exhibit} a wide range of horizontal length { scales}. The most obvious feature would be the meandering of the equatorial jets, as visibly seen in the temperature maps. All models with $\tau_{\rm{rad,top}}=10^4$ and $\tau_{\rm{rad,top}}=10^5\rm\,s$ show more or less meandering of their equatorial jets.  Even smaller scale eddies { exhibiting} high { frequency oscillations are present on} the flanks of the equatorial jet, but they do not play a major role in maintaining the circulation. {\ttt The PV structure near the flanks of the equatorial superrotating jet shown in Figure \ref{closebd.pv} exhibits strong poleward decrease in amplitudes  (in regions between about $ 2^{\circ}$ to $7^{\circ}$ and between $ -2^{\circ}$ to $-7^{\circ}$) at both 80 mbar and 0.93 bar, indicating  violation of  the barotropic stability criteria $\partial^2\overline{u}/\partial y^2<\beta$ (e.g., \citealp{ingersoll1982}). This suggests that the equatorial jet is barotropically unstable in the model with  a rotation period of 2.5 hours, which is likely responsible for the meandering of the jet and small-scale transient eddies at the flanks of the equatorial jet.  } { Whether} barotropic instability would occur in hot-Jupiter atmospheres and its effect on limiting the jet speed have been discussed in { the} literature \citep{menou2009,heng2011,fromang2016,menou2019}. All { our} models with $\tau_{\rm{rad,top}}=10^4$ s strongly violate the barotropic stability { criterion}, and apparently barotropic instability { occurs in these models}. Models with $\tau_{\rm{rad,top}}=10^5$ s also slightly violate the { barotropic stability criterion}. However, { such instabilities may not be a critical factor in limiting the equatorial jet speed in our models;}  otherwise the instability would efficiently remove the kinetic energy associated with the strong wind shear by enforcing the jet profile to be neutral to barotropic stability.

The transition of dynamical regime with respect to rotation impacts the phase-curve amplitude and offset. With decreasing rotation period, the global day-night temperature difference visibly { increases}, and straightforwardly the day-night flux difference is expected to increase. For relatively short rotation period, at low latitude within the Matsuno-Gill pattern, the hottest area is always east of the substellar point due to the strong equatorial superrotating jet and the equatorial Kelvin wave. The off-equatorial anti-cyclonic Rossby gyres are west of the substellar point and are associated hot regions. As the rotation period decreases, the projected area of the Rossby gyres increase and more thermal flux from { west} of sub-stellar point contributes to the disk-integrated flux, while the meridional extent of the equatorial region decreases. The combination of the two reduces or even reverses the phase-curve offset that is typically observed to peak before the secondary eclipse for canonical hot Jupiters. In addition, the off-equatorial Rossby gyres sometimes can be much hotter than the equatorial region and exhibit fluctuations, and there exist hot “tails” west of the Rossby gyres that can extend to the nightside. { We expect that} these features { will} influence the phase curve offset and induce time variability. Synthetic phase curves will be shown in Section \ref{ch.result5}.

\begin{figure}      
\epsscale{1.15}      
\plotone{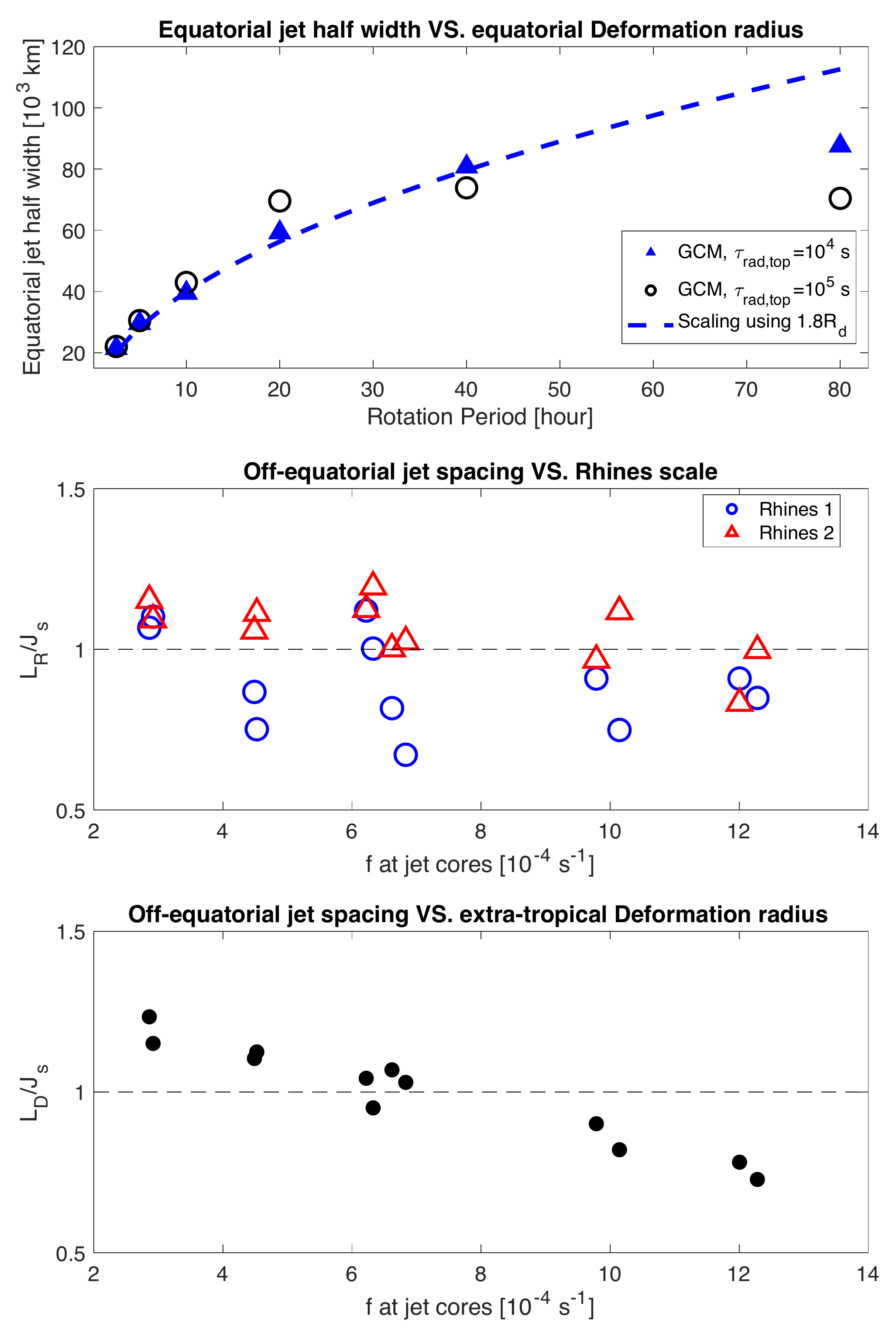}
\caption{Zonal jet spacing and scaling using the Deformation radius and Rhines scale.   \emph{Top panel}: comparisons of the measured equatorial superrotating jet half width as a function of rotation period  to the equatorial deformation radius (scaled up by a factor of 1.8) as a function of rotation period (dashed line) assuming an isothermal atmosphere and $T\sim 1400$ K. Triangles are measurements from models with $\tau_{\rm{rad,top}}=10^4$ s and circles are from those with $\tau_{\rm{rad,top}}=10^5$ s.  \emph{Middle panel}: The ratio  between  the Rhines scale defined using vertically mass-weighted eddy velocity and the off-equatorial jet spacing (circles),  and the ratio  between  the Rhines scale defined using eastward jet velocity and the off-equatorial jet spacing (triangles). These ratios are shown as a function of the Coriolis parameter $f$ at jet cores. The measured jet spacing are based on jet profiles vertically averaged between 1 bar and $10^{-3}$ bar for GCMs with $\tau_{\rm{rad,top}}=10^4$ s and with $10^5$ s.  \emph{Bottom panel}: The ratio between the extra-tropical deformation radius (assuming an isothermal atmosphere and $T\sim 1400$ K) and  the off-equatorial jet spacing.}
\label{closebd.halfwidth}
\end{figure}

We now discuss { the structure} of the zonal jets, the relation to relevant dynamical length scales and { jet-driving} mechanism in our models. We start off with the equatorial superrotating jets. In Section \ref{ch.expectation} we { summarized} the mechanism that drives the equatorial superrotation { on} tidally locked exoplanets, { namely} the interactions { of the standing tropical waves with the mean flow, and this theory predicts that the equatorial jet has a half-width approximately equal to the equatorial deformation radius,} $\sqrt{NHa/\Omega}$ \citep{showman2011}. { We measure the half-widths of the equatorial jets in our GCM simulations to compare them to this scaling theory.} The measured half widths of the equatorial superrotating jets from GCMs scale quite well with our analytic prediction, showing a close correlation with $\Omega^{-1/2}$. The half width of the equatorial jet is defined as the distance between the equator and where the jet profile first reaches a minimum. The jet profiles used here are vertically averaged between 1 bar and $10^{-3}$ bar. In the top panel of Figure \ref{closebd.halfwidth} we show the jet half widths measured from GCMs with $\tau_{\rm{rad,top}}=10^4$ and $10^5$ s and comparisons to the equatorial deformation radius (scaled up by a constant 1.8) as a function of rotation period assuming an isothermal atmosphere with $T\sim 1400$ K. For models with $\tau_{\rm{rad,top}}=10^4$ s, the scaling is particularly good { for} rotation periods $\lesssim 40$ hour. { For} rotation periods $> 40$ hours, the measured jet half-width { flattens} out with increasing rotation period. For models with $\tau_{\rm{rad,top}}=10^5$ s, the { flattening} starts at 20 hours. The mismatch { at} long rotation period is partly because the wave theory and the corresponding scaling \citep{showman2011} is established { on} the equatorial $\beta$ plane, a local approximation that could overestimate the jet width when the waves and wave-mean-flow interactions can take place at mid-high latitudes where the full spherical geometry should be taken into account. In the case of slow rotators, the deformation radius is comparable to the planetary radius, and thus the accuracy of our scaling is affected.  It is common in practice that GCM simulated jets differ from the theoretical scaling by an O(1) constant (e.g., \citealp{schneider2006,schneider2009}). In reality, several factors may also affect this comparison. First, the scaling is based on a { theory} that is linearized about a rest  state, and our measured jet width is based on the fully equilibrated, nonlinear calculations.  Second, the wave structure and wave-mean-flow interactions would be altered by the presence of a strong equatorial jet \citep{tsai2014,hammond2018}.  Third, the nonlinear effects may not be precisely  quantified as the frictional drag. Finally, dynamics of a continuously stratified atmosphere is considerably more complicated than that of the shallow-water system. Given all the complexities, the scaling of the width of the equatorial superrotating jet works surprisingly well in explaining our GCM results and supports a uniform mechanism that drives the equatorial superrotation from slow-rotating regime to rapid-rotating regime.

Now we look at the off-equatorial jets. It { is} widely accepted that the off-equatorial jet spacing in simulations of rapid rotating, thin atmospheres { subject to small-scale turbulent forcing} is typically close to the Rhines scale (see a recent review by \citealp{showman2019review}), a scale over which two-dimensional turbulence interactions becomes anisotropic due to planetary rotation \citep{rhines1975,vallis1993}. Here we compare the measured off-equatorial jet spacing to the Rhines scale in the middle panel of Figure \ref{closebd.halfwidth}, which shows the ratio between Rhines scale and jet spacing as a function of the Coriolis parameter $f$ at the eastward jet core for models with rotation period 10 hours and less and with $\tau_{\rm{rad,top}}=10^4$ and $10^5$ s.   Like the equatorial jet, the off-equatorial jet spacing is measured using a vertically averaged jet profile and is the distance between the eastward jet flanks where the jet profile reaches local minimum. There are two flavors of the Rhines scale which are commonly used in the literature. The first one is expressed using a characteristic eddy velocity  $L_R\sim 2\pi \sqrt{U_{\rm{eddy}}/\beta}$ (e.g., \citealp{rhines1975,schneider2006,chemke2015}) where $U_{\rm{eddy}}$ used here is vertically mass-weighted eddy volocity between 1 and $10^{-3}$ bar following \cite{schneider2009}. The other one is expressed using a characteristic jet velocity $L_R\sim \pi \sqrt{2U_{\rm{jet}}/\beta}$ (e.g., \citealp{williams1978,lian2008}) where $U_{\rm{jet}}$ here is the vertically averaged eastward jet core speed.\footnote{{\tt The interpretation of these two versions of the Rhines scale differ.  The first comes from simple turbulence scaling arguments.  The second can be obtained from imagining a series of zonally symmetric zonal jets that correspond to a potential-vorticity (PV) staircase, with nearly constant strips of PV joined together by sharp PV gradients.  This situation naturally corresponds to jets whose meridional spacings and zonal-mean zonal wind speeds are related by the second Rhines scale.  See for example \cite{scott2012}.}} The eddy Rhines scale ratios are { represented} as circles in the middle panel of Figure \ref{closebd.halfwidth}, and { those scaled using zonal-mean jet speed are plotted as} triangles. These ratios are scattered around one within maximum deviation of about 1.4  over a wide range of $f$. Like many previous investigations, the off-equatorial jet spacing scales well with the Rhines scale. The Rhines scale based on both the eddy and jet velocities are equally well fitted to our jet spacing. This is perhaps because at mid-high latitudes of our rapid rotators,  kinetic energy associated with eddies is comparable to that associated with the zonal jets. It is interesting to note that kinetic energy near the equator is dominated by the strong equatorial superrotating jet, which contains more than 95\% of the total kinetic energy, and this is the case for all models with different rotation period and with $\tau_{\rm{rad,top}}=10^4$ and $10^5$ s. 

The extratropical deformation radius $L_D \sim 2\pi c/f$ where $c$ is an appropriate gravity phase speed is a relevant energy injection scale, especially for jets that are driven by baroclinic instability. The bottom panel in Figure \ref{closebd.halfwidth} shows the ratio between the extratropical deformation radius (with $c\sim 2NH$ assuming isothermal atmosphere with $T\sim 1400$ K) and the jet spacing as a function of $f$. Likewise the deformation radius also exhibits { relatively good agreement} to the jet spacing, with a maximum deviation { of} about 1.3.   There is little scale separation between the Rhines scale and the deformation radius shown in Figure \ref{closebd.halfwidth}, suggesting { that} nonlinear eddy-eddy interactions { are weak} in our fast rotating models, similar to { the} extratropical regime { in Earth's} atmosphere (e.g., \citealp{held1996,schneider2006}). The zonal jets are maintained mainly by interactions with eddies generated from baroclinic instability and the day-night forcing, and nonlinear eddy-eddy interactions { appear not to be} essential in determining the jet structure. { There} is an emerging trend that $L_D/J_s$ is larger than 1 at small $f$ and systematically decreases to below 1 at large $f$. This trend reflects a tendency of scale separation as a function of local Coriolis parameter, which has been documented in previous studies (e.g., \citealp{chemke2015}).

 Comparison of the zonal jets between models with $\tau_{\rm{rad,top}}=10^4$ and $\tau_{\rm{rad,top}}=10^5$ s (Figure \ref{closebd.uzonalav}) is interesting.  The spacing and strength of the off-equatorial zonal jets, as well as the associated zonal-mean meridional temperature gradients are quantitatively  similar between models with $\tau_{\rm{rad,top}}=10^4$ and $\tau_{\rm{rad,top}}=10^5$ s. { In contrast,} for the equatorial jets, despite their  similar jet width, the jet speed with $\tau_{\rm{rad,top}}=10^5$ s is almost half of that with $\tau_{\rm{rad,top}}=10^4$ s. The fact that given a fixed $\tau_{\rm{rad,top}}$ the equatorial jet speed is almost the same among models with different rotation rate is also worth noticing.  { Recall} that frictional drag is absent { at pressures less than} 10 bars in models shown in Figure \ref{closebd.uzonalav}. The above comparisons indicate that, in the relatively short-$\tau_{\rm{rad,top}}$ regime, scaling of the off-equatorial jet strength { with rotation rate} differs from that for the equatorial jet. No theory currently exists to predict the equatorial jet speed of tidally locked planets, and theories predicting off-equatorial jet speed often apply only { under} conditions of { zonally symmetric,} equator-to-pole forcing. \cite{zhang2017} showed that the scaling of eddy velocity using the theory of  \cite{komacek2016} provides a good match to the equatorial jet speed, indicating the intimate relation of eddy fluxes and the equatorial jet speed.  These issues are challenging and we do not attempt to resolve them here.

\begin{figure*}      
\epsscale{1.15}      
\plotone{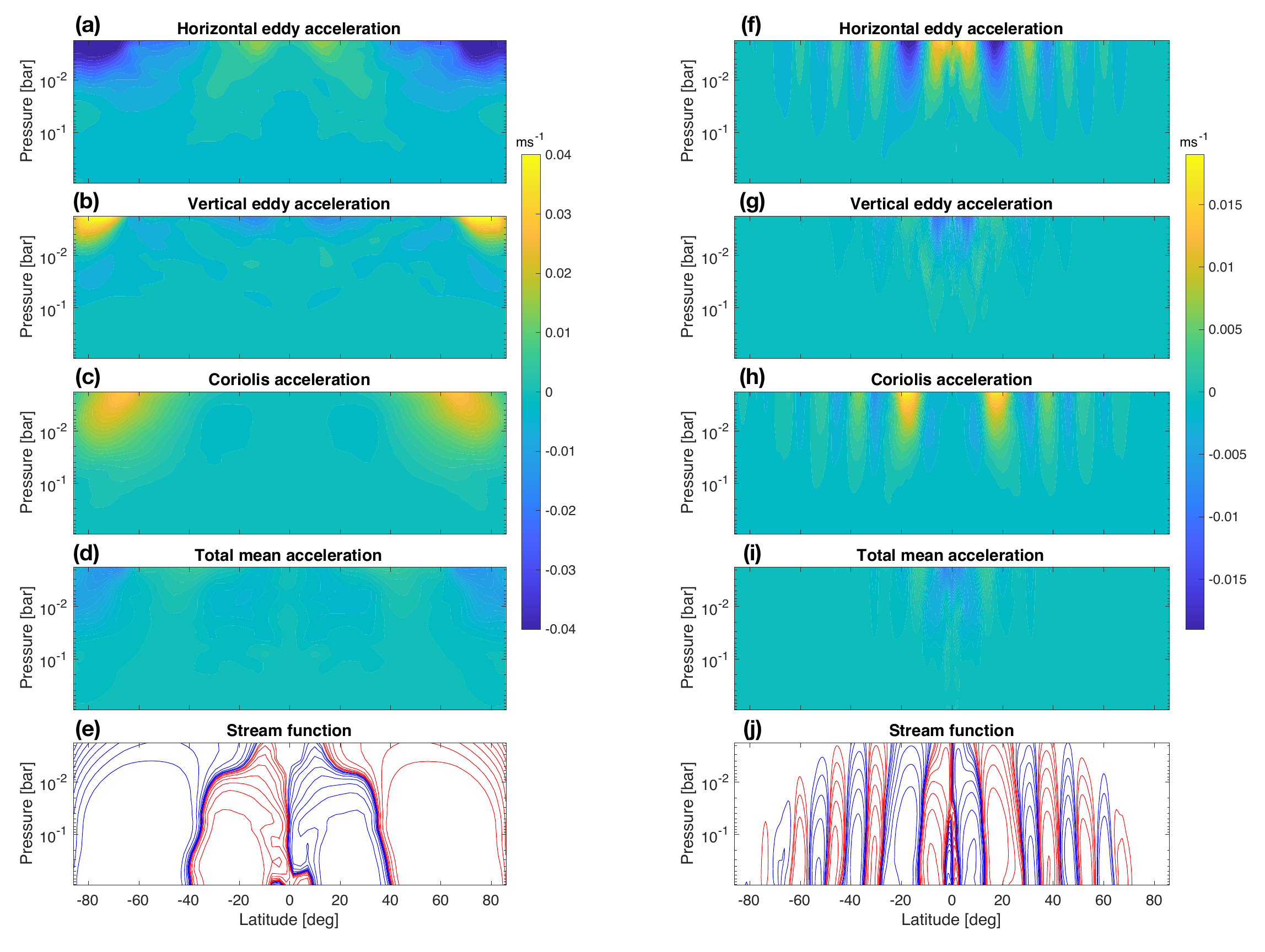}
\caption{{\ttt Zonal-mean diagnostics of the model with a rotation period of 40 hours and $\tau_{\rm rad,top}=10^4$ s on the left column and the model with a rotation period of 2.5 hours and $\tau_{\rm rad,top}=10^4$ s on the right column.  {\bf (a)} shows  the  horizontal eddy acceleration $\left({\rm the ~term~} - \frac{1}{a\cos^2\phi}\frac{\partial(\overline{u'v'}\cos^2\phi)}{\partial \phi} {\rm ~in ~Equation ~(\ref{eq.diagnose})}\right)$  on the zonal-mean zonal wind; {\bf (b)} shows the vertical eddy acceleration $\left(- \frac{\partial(\overline{u'\omega'})}{\partial p}\right)$ on the zonal-mean zonal wind; {\bf (c)} shows the Coriolis acceleration associated with the zonal-mean meridional circulation ($\overline{v}f$); {\bf (d)} shows the acceleration by zonal-mean momentum transport  $\left(- \frac{\overline{v}}{a\cos\phi}\frac{\partial(\overline{u}\cos\phi)}{\partial \phi}-\overline{\omega}\frac{\partial\overline{u}}{\partial p}\right)$; and finally {\bf (e)} shows the zonal-mean stream function. In the bottom panels, red lines are clockwise with log-spaced contours between 3$\times 10^8$ and 3$\times 10^{10}\;{\rm kg\;s^{-1}}$, and blue lines are counterclockwise with log-spaced contours between $-3\times 10^8$ and $-3\times 10^{10}\;{\rm kg\;s^{-1}}$. Panels {\bf (f)} to {\bf  (j)} are similar but for the model with a rotation period of 2.5 hours and $\tau_{\rm rad,top}=10^4$ s.}
}
\label{closebd.diagnose}
\end{figure*}

{\ttt Finally, we briefly present diagnostics that further illustrate the role of eddies in maintaining the zonal-mean circulation and the transition from slow to rapid rotation. The zonal-mean zonal momentum equation of primitive equations in pressure coordinates can be written as 
\begin{equation}
\begin{split}
\frac{\partial \overline{u}}{\partial t}=\overline{v}f - \frac{\overline{v}}{a\cos\phi}\frac{\partial(\overline{u}\cos\phi)}{\partial \phi}-\overline{\omega}\frac{\partial\overline{u}}{\partial p}  \\ - \frac{1}{a\cos^2\phi}\frac{\partial(\overline{u'v'}\cos^2\phi)}{\partial \phi} - \frac{\partial(\overline{u'\omega'})}{\partial p},
\end{split}
\label{eq.diagnose}
\end{equation}
where $f$ is the Coriolis parameter,  $\phi$ is latitude, $\overline{u},\overline{v}$ and $\overline{\omega}$ are zonal-mean zonal, meridional and vertical velocities in pressure coordinates, and $u',v'$ and $\omega'$ are their deviations from the zonal average, respectively. The terms on the right-hand side represent Coriolis acceleration associated with the mean moridional flow, meridional momentum advection by the mean flow, vertical momentum advection by the mean flow, horizontal convergence of  eddy momentum and vertical convergence of eddy momentum.}

{\ttt Accelerations on the zonal-mean zonal wind by various terms on the right-hand side of Equation (\ref{eq.diagnose}) for a slowly rotating model with  a rotation period of 40 hours and $\tau_{\rm rad,top}=10^4$ s are shown on the left column of Figure \ref{closebd.diagnose}. In agreement with previous hot Jupiter studies (e.g., \citealp{showman2011,showman2015,mayne2017}), the standing Matsuno-Gill  pattern driven by the  day-night thermal forcing generates horizontal eddy convergence of momentum that is responsible for maintaining the equatorial superrotation, and is mainly balanced by vertical eddy convergence at low latitudes. The acceleration is the strongest near the flanks of the equatorial jet rather than  at the center of the jet, which  is slightly different to that in the spin-up phase (see, for example, \citealp{showman2011}, \citealp{showman2015}). This is presumably caused by  modification of the standing wave pattern  by the strong equatorial jet, and has been similarly shown in GCM diagnostics of \cite{mayne2017}. At high latitudes, the Coriolis acceleration associated with the mean meridional circulation becomes important in the balance of  zonal  angular momentum. Overall, the momentum advection by mean flows are not dominant at all latitudes and pressures. 
}

{\ttt Diagnostics of the rapidly rotating model with a rotation period of 2.5 hours and $\tau_{\rm rad,top}=10^4$ s are shown on the right column of Figure \ref{closebd.diagnose}. Near the equator (within $\pm 10^{\circ}$), despite being confined at much lower latitudes, excitation of the Matsuno-Gill pattern by the day-night forcing is responsible for maintaining the equatorial superrotation, similar to that in the slowly rotating model. At mid-high latitudes (poleward of $\pm 10^{\circ}$), the maintenance of subtropical eastward jets are due to the horizontal eddy convergence of momentum by the baroclinic eddies (see positive accelerations by horizontal eddy transport at about $\pm 30^{\circ}$, $\pm 45^{\circ}$ and $\pm 60^{\circ}$ in panel (f), which correlates well to the eastward jets shown in Figure \ref{closebd.uzonalav} and \ref{closebd.pv}). However,  the major balance in the zonal momentum at mid-high latitudes is between the Coriolis acceleration and horizontal eddy convergence (see a comparison between panel (f) and (h) in Figure \ref{closebd.diagnose}), namely
\begin{equation}
    f\overline{v}\approx \frac{1}{a\cos^2\phi}\frac{\partial(\overline{u'v'}\cos^2\phi)}{\partial \phi}.
    \label{eq.balance}
\end{equation}
This is consistent with the fact that dynamics is quasi-geostrophic at mid-high latitudes, and the above balance in the statistically steady state is well-known in this regime (e.g., \citealp{schneider2006review,vallis2006,holton2012}). 
}

{\ttt The zonal-mean stream function $\overline{\psi}$, defined by $\overline{v}=g(2\pi a\cos\phi)^{-1}\partial \overline{\psi}/\partial p$ and $\overline{\omega}=-g(2\pi a^2\cos\phi)^{-1}\partial \overline{\psi}/\partial \phi$, where $a$ is planetary radius, is shown in the bottom panels of Figure \ref{closebd.diagnose} for the slow rotating model with a rotation period of 40 hours on the left and the rapid rotating model with a rotation period of 2.5 hours on the right. Red lines are clockwise and blue lines are counterclockwise. There are {\bbb mainly} two zonal-mean meridional circulation cells {\bbb above 0.5 bar} in each hemisphere in the slowly rotating case. They are separated at about $\pm 40^{\circ}$, which is consistent with the change of eddy acceleration in the  meridional direction (panel  (a) and (b)). The air associated with the mean meridional circulation descends at the equator and ascends near $\pm 40^{\circ}$,  indicating a thermally indirect zonal-mean meridional circulation near the equator. {\bbb Interestingly, there are two thermally direct circulation cells within about $\pm10^{\circ}$ at a deeper pressure than the thermally indirect cells.}  The zonal-mean stream function of the rapid rotator (panel (j)) exhibits up to 9 cells in each hemisphere, which is not surprising given the multiple zonal banded structure in the simulation. At low latitudes, air associated with the mean meridional circulation ascends at about $\pm 10^{\circ}$ and descends {\bbb near} the equator, and is similarly a thermally indirect circulation. {\bbb Similar to the slow rotating case, there are small thermally direct cells within about $\pm 3^{\circ}$ at a pressure deeper than 50 mbar.}   At mid-high latitudes, as indicated by the balance in Equation (\ref{eq.balance}), the mean meridional  circulation is eddy-driven and the strength is constrained by magnitude of the  eddy-momentum convergence. Indeed, the position and strength of the off-equatorial cells correlate well to the horizontal eddy acceleration shown in panel (f). The eddy-driven circulation at mid-high latitudes is similar to the picture that has been proposed for the off-equatorial jets on  Jupiter and Saturn near the cloud level \citep{showman2007,delgenio2007,schneider2009}.
}

\subsection{Circulation with long radiative timescale}
\label{ch.result2}

Models with long radiative timescales ($\tau_{\rm{rad, top}}=10^6$ and $10^7$ s) show { qualitatively} different horizontal structure { from} those with short radiative timescales, { analogous to the behavior} found in \cite{perezbecker2013} and \cite{komacek2016}. As straightforwardly seen in Figure \ref{closebd.temp30} and \ref{closebd.temp23}, given any rotation period, the day-night temperature differences at low pressure are much smaller { in models with long radiative timescales} than those with short radiative timescales. In slow rotators with 40 and 80 hour rotation period, the horizontal temperature structure is rather zonal  even at low pressure.  Surprisingly, in models with rotation period 20 hours and less, a global-scale wave pattern can still be visibly seen at 7 and 80 mbar, { albeit} with much weaker winds and temperature perturbations than those with shorter $\tau_{\rm{rad,top}}$. For models with 5 and 2.5 hour rotation, the wave patterns are still vaguely seen even { at} 1.3 bar.  For rapid rotators, in contrast to those with $\tau_{\rm{rad, top}}=10^4$ and $10^5$ s, the circulation with long $\tau_{\rm{rad, top}}$ is overall  characterized by meridionally broad wave structures,  showing no obvious multiple zonal banded structures. It is possible that with long radiative timescale, global scale waves are the major response to the day-night forcing, and the waves significantly redistribute heat around the globe, eliminating the large equator-to-pole temperature difference. As a result, instability and small-scale { eddies} do not easily occur in these models.

Equatorial superrotation { exists} in all models with $\tau_{\rm{rad, top}}=10^6\rm\,s$, { although} with weaker wind speeds than those with $\tau_{\rm{rad, top}}=10^5\rm\,s$ (see Figure \ref{closebd.uzonalav}). { In rapid rotators,} the depth of the equatorial jets { at $\tau_{\rm rad, top}=10^6\rm\,s$} is shallower than that with $\tau_{\rm{rad, top}}=10^4$ and $10^5\rm\,s$ by more than a scale height; the off-equatorial zonal flows are much weaker, and their meridional shapes are less well developed than those with $\tau_{\rm{rad, top}}=10^4$ and $10^5\rm\,s$. These features are not surprising as in general the magnitude of eddies gets weaker with longer $\tau_{\rm{rad, top}}$, and the zonal jets driven by them { become} weaker. Interestingly, for models with 40 and 80 hour rotation, although there is equatorial superrotation, the jet profile actually peaks { off the} equator, indicating a stronger role of meridional circulation in shaping the overall jet structure than that with shorter $\tau_{\rm{rad, top}}$. For models with $\tau_{\rm{rad, top}}=10^7$ s, slow rotators exhibit qualitative similar jet structures to those with $\tau_{\rm{rad, top}}=10^6$ s. However models with 20-hour rotation period and less have little or no equatorial superrotation, and their overall circulation is yet much weaker than those with $\tau_{\rm{rad, top}}=10^6$ s.

\begin{figure}      
\epsscale{1.2}      
\plotone{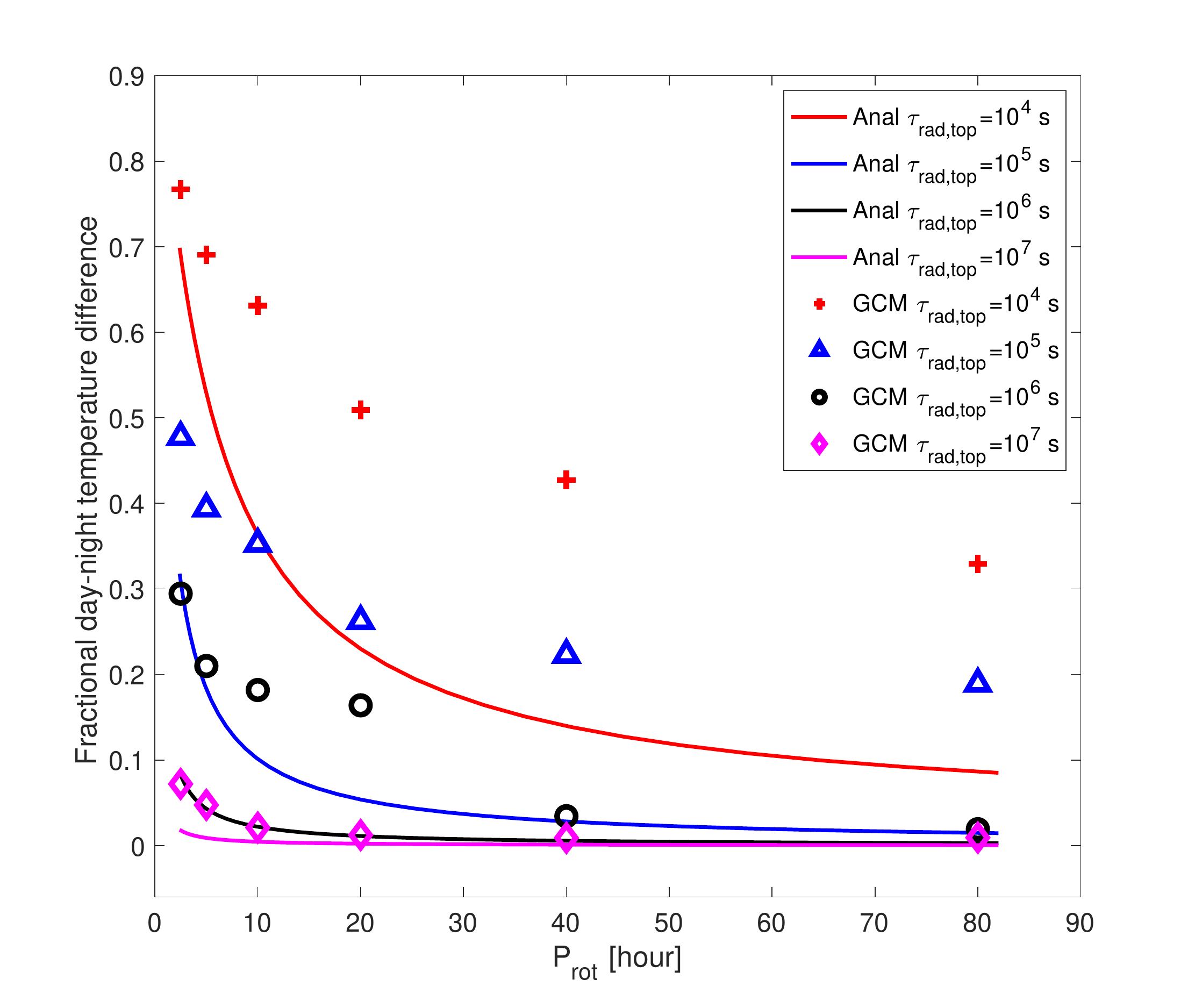}
\caption{Scatter points are the day-night temperature differences measured by the pressure dependent quantity $A = \Delta T/ \Delta T_{\rm{eq}}$ at 80 mbar (as defined in \citealp{komacek2016}) for drag-free models with different rotation period and radiative timescale $\tau_{\rm{rad, top}}$. Solid curves are predictions using the scaling theory of \cite{komacek2016} and \cite{zhang2017}. $A\sim 0$ represents homogeneous longitudinal temperature distribution and $A\sim 1$ represents a temperature distribution close to radiative equilibrium.  }
\label{closebd.daynightA}
\end{figure}

We summarise results over the entire grid by discussing the fractional day-night temperature differences. The  day-night temperature difference is mainly regulated by wave adjustment. This process competes with  radiative damping,   frictional drag damping and planetary rotation \citep{perezbecker2013, komacek2016}.  The fractional day-night temperature differences quantified using the nondimensional quantity $A = \Delta T/ \Delta T_{\rm{eq}}$ (as defined in \citealp{komacek2016}) at 80 mbar for models with different rotation periods and radiative timescales  are shown in Figure \ref{closebd.daynightA} as scatter points. Predictions using  the scaling theory of \cite{komacek2016} and \cite{zhang2017} are plotted as solid lines as a comparison. $A\sim 0$ represents homogeneous longitudinal temperature distribution and $A\sim 1$ represents a temperature distribution close to radiative equilibrium. Overall, the day-night temperature difference increases with decreasing rotation period and decreasing radiative timescale. Long radiative timescale results in negligible day-night temperature difference despite short rotation period. 

The scaling theory captures the overall trends of the GCM results, consistent with the findings in \cite{komacek2016} and \cite{zhang2017}, { but} systematically predicts lower fractional day-night temperature differences than that from our GCMs { (Figure~\ref{closebd.daynightA})}.  As shown in \cite{komacek2016}, the analytic theory works best in the strong-drag  regime  which typically has a laminar day-night flow, but relatively less well in the drag-free regime where the strong equatorial jet { is} driven. { The \cite{komacek2016} scaling theory has no explicit representation of the equatorial jet or the Matsuno-Gill standing wave pattern; the theory assumes that the wind flows straight from day to night and that the temperature structure comprises a simple day-night difference, which are assumptions that (for a given wind speed and day-night temperature difference) maximize the day-night heat transport.  In reality, Figures~\ref{closebd.temp30} and \ref{closebd.temp23} make clear that winds outside the equatorial jet often follow nearly parallel to isotherms (this is true not only for very rapid rotators but also to some degree even for periods of 10, 20, and 40 hours).  Only the component of wind {\it perpendicular} to isotherms will advect temperature, and this component is weaker than the total wind amplitude. This phenomenon suggests that---for a given root-mean-square wind speed and day-night temperature difference---the actual day-night heat transport will be smaller than predicted by the theory, and therefore the equilibrated day-night temperature difference will be larger than predicted by the theory.  It may be possible to improve the theory to account for some of these effects, a task we leave for the future.}

 
\subsection{The role of frictional drag}
\label{ch.result4}

The thermal ionization fraction could be significant in hot atmospheres of extremely close-in gas giants in the sense that the atmospheric flows can be strongly coupled to the intrinsic magnetic field and experience significant magnetohydrodynamic drag. {\ttt Some GCMs parameterize this dissipation as simple frictional drag  and explored its consequences on the flows (e.g., \citealp{perna2010, perna2012, komacek2016}). There have been self-consistent  magnetohydrodynamic models  but with simplified thermal forcing schemes (e.g., \citealp{rogers2014komacek, rogers2017}). Both models predict slower winds due to the Lorentz force. }  Other possible dissipation { mechanisms include} turbulent dissipation (e.g., \citealp{li2010}).  To crudely represent the effects of magnetic or turbulent effects on reducing the wind speed and shaping the circulation pattern, we include a pressure-independent frictional drag in a set of model with 2.5-hour rotation period and $\tau_{\rm{rad,top}}=10^4$ s, and with two different drag timescale $\tau_{\rm{drag}}=10^6$ and $10^4$ s. 

\begin{figure*}     
\epsscale{1.1}
 \plotone{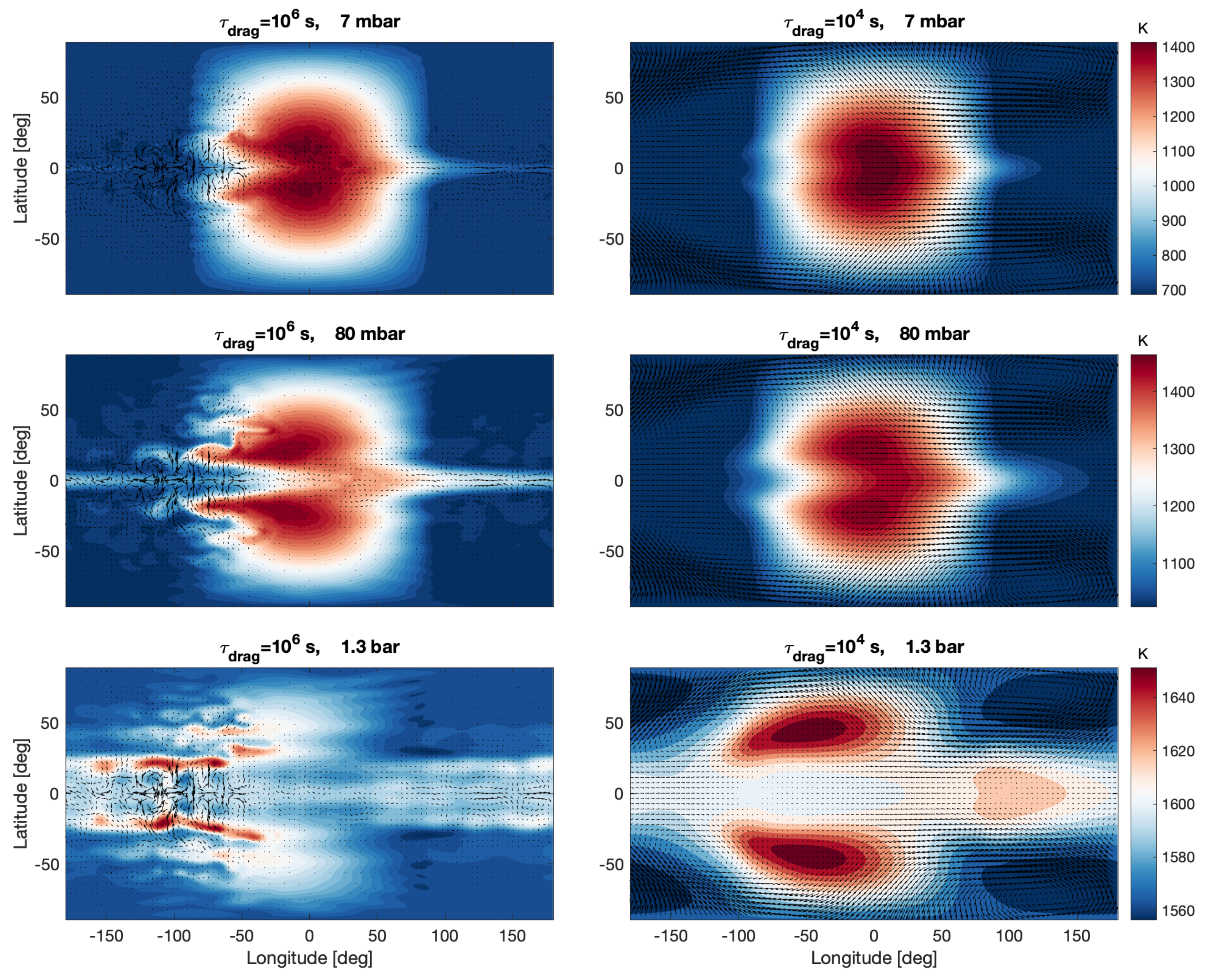}
\caption{Snapshots of horizontal temperature map with { superposed} wind vectors at 7 mbar in the first row, 80 mbar in the second row and 1.3 bar in the third row for models with 2.5-hour rotation period, $\tau_{\rm{rad,top}}=10^4$ s, and different frictional drag timescale  $\tau_{\rm{drag}}=10^6\rm\,s$ (left column)  and $10^4\rm\,s$ (right column). }
\label{closebd.drag}
\end{figure*}

Figure \ref{closebd.drag} shows horizontal temperature maps with overplotted winds at three different pressure levels for models with  $\tau_{\rm{drag}}=10^6\rm\,s$ on the left, and with $\tau_{\rm{drag}}=10^4\rm\,s$ on the right.  The overall circulation pattern of the model with $\tau_{\rm{drag}}=10^6\rm\,s$ at relatively low pressure resembles qualitatively that of the drag-free model, showing a strong, robust equatorial superrotation, characteristic Matsuno-Gill wave patterns at low latitudes and nearly geostrophic flow at high latitudes. The magnitude of wind speed is lower than { in our} drag-free { models} due to the drag. { Small-scale} vortices and waves are visibly seen at both low and high latitudes, which are similarily caused by baroclinic and barotropic instability. A quantitative difference is that there is no obvious  zonal banding even at high pressure (1.3 bar shown in Figure \ref{closebd.drag}) despite the existence of eddies. The drag timescale is likely { shorter than the dynamical timescale on which eddies can accelerate the zonal jets and over which turbulence can reorganize into a zonalized configuration.  Thus, when the drag timescale is sufficiently short, it plays a critical role in preventing off-equatorial jet formation.  These results are consistent with those of \cite{perna2010}, \cite{showman2013b}, and others, who described the role drag can play in suppressing jet formation in the deeper layers of the troposphere.} 

In the model with $\tau_{\rm{drag}}=10^4$ s, the drag timescale is much closer to the rotation timescale, and thus the overall horizontal force balance { transitions} from the geostrophic regime to a three-way balance regime. At { mid-to-high} latitude, this can be recognized { due to the fact} that the wind vectors do not necessarily follow isotherms { on isobaric surfaces} (right column in Figure \ref{closebd.drag}). The equatorial jet is suppressed. The off-equatorial Rossby waves are still seen but much weakened. The pattern is again quite similar to predictions made by using linearized analytic models.  With an even stronger drag $\tau_{\rm{drag}}= 10^3$ s (not shown),  the circulation pattern is dominated by the pure day-to-night flow, with a weak flow speed and a temperature pattern { nearly in radiative equilibrium}. In this case the drag timescale  is shorter than any other relevant dynamical timescales (especially the rotation timescale), and  the force balance is simply between pressure gradient and drag. { The weak wind speeds allows only relatively weak day-night heat transport, helping to preserve the temperature structure close to radiative equilibrium}.   The existence of strong drag has important consequence { for} the thermal phase curve due to the influence on the horizontal temperature structure.  The phase curve { at} different wavelengths can { provide} a diagnostic tool { to discriminate} the extent to which the drag shapes the circulation. { Still, Lorentz forces will have a far more complex pattern than allowed by our simple Rayleigh-drag scheme, so these strong-drag models should be viewed more as experiments to understand how the circulation responds to idealized drag of various strengths, and not as predictions for how actual atmospheres affected by Lorentz forces or strong turbulent mixing will behave.}

\subsection{Synthetic phase curves}
\label{ch.result5}

\begin{figure*}      
\epsscale{1.1}      
\plotone{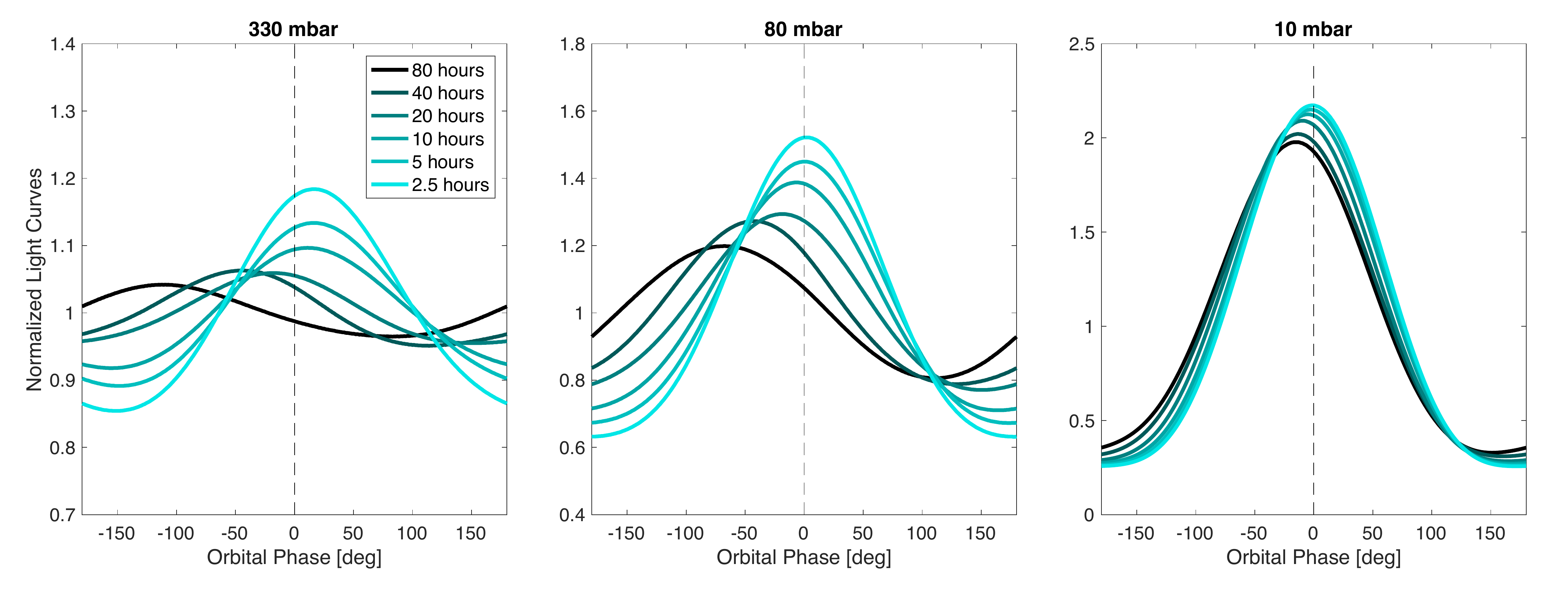}
\caption{Normalized  synthetic thermal phase curves sampling different pressure levels for drag-free models with different rotation period from 80 to 2.5 hours and a short radiative timescale $\tau_{\rm{rad,top}}=10^4$ s. These phase curves are obtained by rotating the modeled atmosphere using time-mean temperature structures. { Although phase curves peak before secondary eclipse in the canonical hot Jupiter regime (black and dark blue curves), the maximum flux tends to occur right at secondary eclipse for very rapid rotators (light blue curves).  This phenomenon can help explain the absence of such offsets in phase curve observations of WD+BD systems \citep[e.g.,][]{casewell2015}.}}
\label{closebd.phasecurve}
\end{figure*}

\begin{figure*}      
\epsscale{1.2}      
\plotone{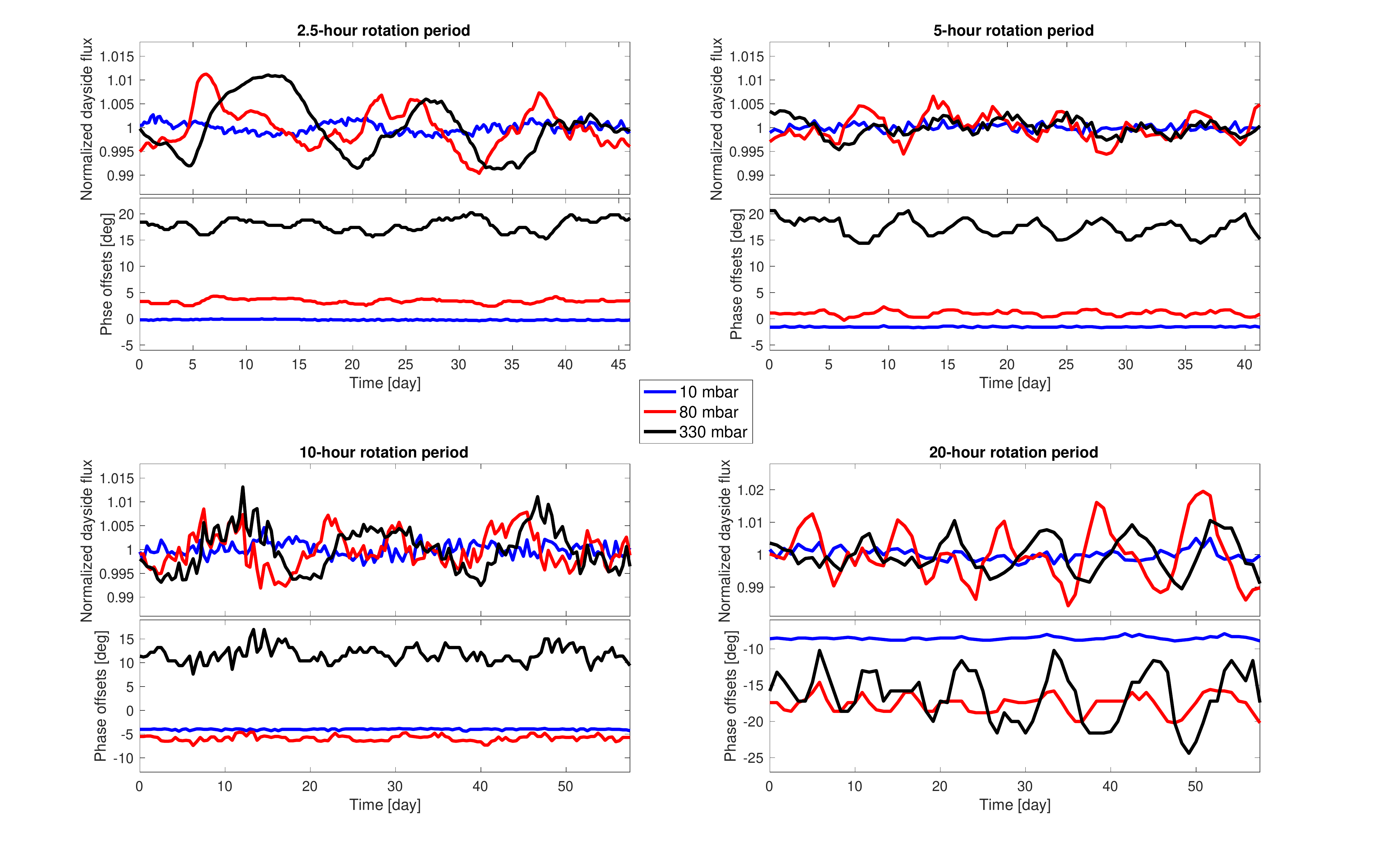}
\caption{These panels show time variability of the synthetic light curves in terms of the dayside flux at the secondary eclipse and the phase offsets of the maximum peak flux for four models with different rotation period as indicated by the titles above each panel. In each case, we show evolution of the normalized dayside flux (normalized around the time-mean value) and phase offsets at three different pressure levels of 10~mbar, 80~mbar and 330~mbar. These models are drag-free { at pressures less than} 10 bars. }
\label{closebd.fluxvariability}
\end{figure*}

\begin{figure*}      
\epsscale{1.1}      
\plotone{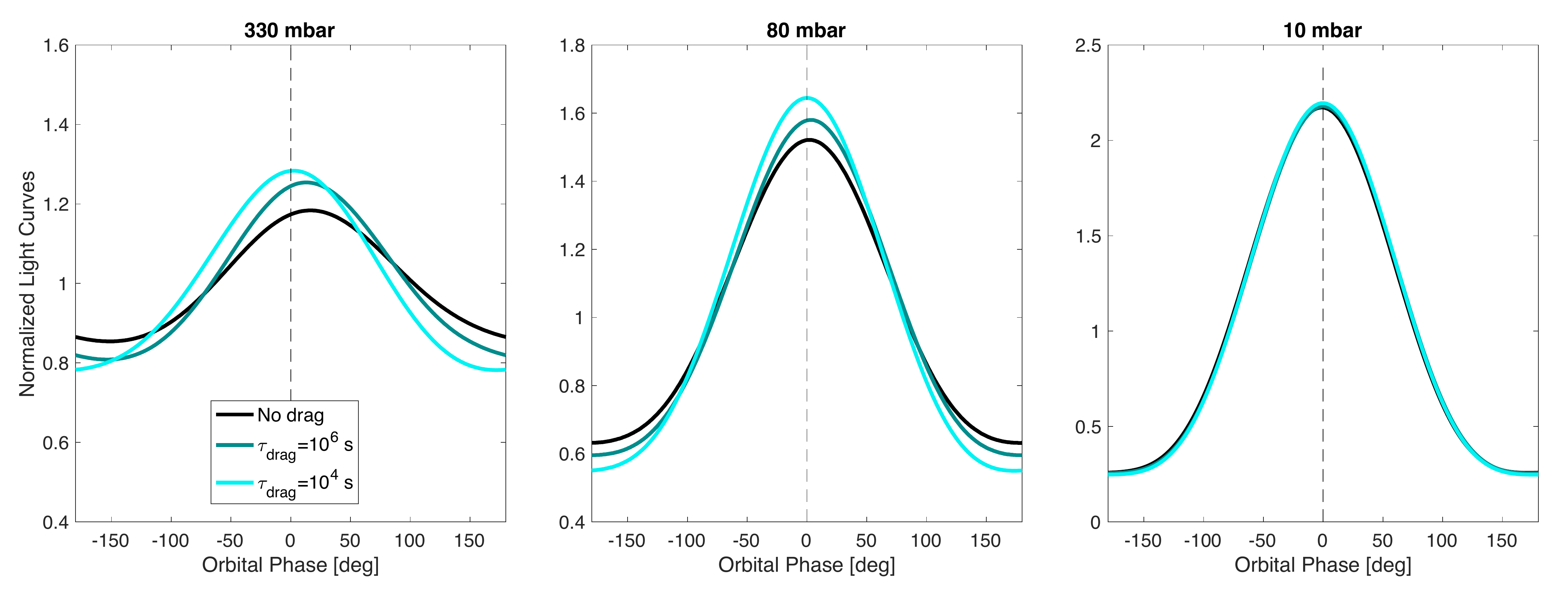}
\caption{Normalized synthetic thermal phase curves  sampling different pressure levels for models with a rotation period 2.5 hours, a short radiative timescale $\tau_{\rm{rad,top}}=10^4$ s  and different frictional drag (no drag, $\tau_{\rm{drag}}=10^6$ and $10^4\rm\,s$).}
\label{closebd.phasecurvedrag}
\end{figure*}

The change of global temperature structure with decreasing rotation period has major implications { for} thermal phase curves.   We  show normalized synthetic thermal phase curves  in Figure \ref{closebd.phasecurve}  sampled from temperature at different pressures for models with different rotation period, a short radiative timescale $\tau_{\rm{rad,top}}=10^4$ s and without frictional drag, using the following relation \citep{zhang2017}:
\begin{equation}
F(\delta, p) = \int_{\pi/2 - \delta}^{3\pi /2-\delta} d\lambda \int_{-\pi/2}^{\pi/2} \sigma T^4(\lambda, \phi, p) R^2_p \cos^2\phi \cos\lambda d\phi .
\end{equation}
Here we assume that the orbit of the BD is aligned with the line of sight of the observer {\tt (i.e., we assume the brown dwarf is a transiting object)}, which maximizes the day-night flux contrast. The temperature structure used for phase curves shown in Figure \ref{closebd.phasecurve} are time-averaged and static as the phase rotates.  A few features are obvious. First, at the same pressure level, the amplitude of phase curves increases with decreasing rotation period, consistent with aforementioned increasing day-night temperature contrast with decreasing rotation period. Second, with the same rotation period, the phase curve amplitude is generally greater {\tt for phase curves sampling lower pressure}, due to the {\tt larger day-night temperature differences that tend to occur at lower pressure relative to higher pressure}. Finally, for slow rotators, the thermal fluxes sampling several pressure levels shown in Figure \ref{closebd.phasecurve} all peak before the secondary eclipse which is attributed to the eastward-shifted equatorial hot spot. As rotation period decreases, the flux peak moves closer to the secondary eclipse. At low pressures (80 and 7 mbar), the thermal flux peak is aligned with the secondary flux for rotation period less than 5 hours. At slightly higher pressure (231 mbar), the flux  peaks after the secondary eclipse for rotation period less than 10 hours.  As we have seen from the horizontal map in Figure \ref{closebd.temp30} and \ref{closebd.temp23},  for faster rotators, the westward-shifted hot areas associated with the Rossby waves move closer to the equator, contributing more to the total flux. Meanwhile, the eastward-shifted hot spot at the equator shrinks, {\tt contributing} less to {\tt the total} flux. The two effects tend to compensate each other, resulting in almost no shift between peak flux and the secondary eclipse at low pressures. {\ttt Note that similar effects on the phase curve has been shown in  \cite{carone2020} who simulated the rapidly rotating hot Jupiter WASP-43b with a rotation of 0.8 day. }

{\tt These results---showing that strong rotation suppresses the offset of flux peak from secondary eclipse---provides an explanation for the absence of a flux-peak offset in many phase curves of BD+WD systems \citep[e.g.,][]{casewell2015}.} Moreover, {\tt the possibility that in some cases the westward-shifted contribution can slightly dominate may explain why} phase curves of some BD+WD systems appear to have a ``westward-shifted hot spot" (for example, the Kepler broad-band phase curve for WD 1202-024 \citealp{rappaport2017}). 


The contribution of {\tt spatial} temperature variability causes  time variability of  the synthetic phase curves. In typical hot-Jupiter simulations, such variability is typically small, around 1\% level \citep{showman2009, fromang2016,komacek2019}, and here we concentrate on variability of the rapid rotating models. Figure \ref{closebd.fluxvariability} {\tt shows} the normalized day-side flux and  phase offsets as a function of time for models with four different rotation period but all with $\tau_{\rm{rad,top}}=10^4$ s and drag-free {\tt at pressures less than} 10 bars. In each case, the upper panel shows dayside flux at the secondary eclipse (normalized by the time-mean value) as a function of time for three pressure levels, and the lower panel shows the phase offsets of the peak thermal flux relative to the secondary eclipse.   In general, the dayside flux emitted from 10 mbar varies less than 1\%, {\tt and} the phase-curve offsets {\tt show almost no variation} for {\tt all rotation periods}. {\tt At} deeper levels of 80 and 330 mbar, the peak-to-peak dayside flux variation  can reach up to about 2.5\% or slightly more in cases with 2.5-hour, 10-hour and 20-hour rotation period, and less than 1\% in the 5-hour case. The phase-curve offsets at 80 mbar can {\tt reach} a few {\tt degrees}, and {\tt those} at 330 mbar are even larger, which in the case with 20-hour rotation period can reach more than $10^{\circ}$ peak-to-peak variation. The typical variability timescale ranges from {\tt several} to more than 10 days. Interestingly, the dayside flux variability generally {\tt exhibits} obvious phase offsets {\tt at} 80 mbar {\tt relative to} 330 mbar {\tt for} all rotation {\tt periods}. The off-equatorial propagating eddies are mainly responsible for the flux and phase-offset variability. To prove this, taking the 2.5-hour case as an example, we have compared synthetic phase curves from the full model, the model with equatorial region (within $\pm 10^{\circ}$ latitudes) replaced with time-mean temperature, the model with off-equatorial region (within $\pm [15^{\circ} ~ 30^{\circ}]$ latitudes) replaced with time-mean temperature, and finally the model with high-latitude region (poleward of $\pm 30^{\circ}$ latitudes) replaced with time-mean temperature. Phase curves from {\tt models} with {\tt the} equatorial and high-latitude {\tt regions} replaced show very similar results to the full model, but phase {\tt curves} from the model with only the mid-latitude region replaced show much less variation as well as a dissimilar evolutionary shape of the light curve.  The phase {\tt difference} of the dayside flux {\tt at} different pressure {\tt levels} is caused by the horizontal phase difference of the propagating waves at different pressure. 

On the other hand, if the circulation suffers from a sufficiently strong drag, the phase-curve amplitude is generally larger than that of drag-free models and there is little phase offset due to the suppression of strong equatorial jet and off-equatorial eddies. Comparison of phase curves sampling different pressures for models with 2.5-hour rotation period and various drag strength (no drag, $\tau_{\rm{drag}}=10^6$ and $10^4$ s) are shown in Figure \ref{closebd.phasecurvedrag}. The effect of drag is small at 10 mbar, but becomes increasingly important at larger pressures. In particular, {\tt for phase curves sampling} 330~mbar, the day-night flux amplitude increases and the phase offset {\tt decreases} with deceasing drag timescale. The dayside flux variability from the model with $\tau_{\rm{drag}}=10^6$ s is much less than 1\%, and the phase curve offset variation is negligible. The drag timescale is comparable or shorter than the timescale that takes the waves in the drag-free model to propagate across the day-night distance, which apparently helps to suppress the temperature variation caused by these waves.  Future phase-curve observations probing both {\tt low and high} pressures would be helpful to disentangle the role of drag in shaping the circulation in these atmospheres.

\section{Discussion}
\label{ch.discuss}

This study made one of the first efforts to understand the atmospheric circulation relevant to brown dwarfs in extremely tight orbits around white dwarfs, and {\tt thoroughly} investigate the role of rotation in shaping the atmospheric circulation of tidally locked planetary companions.  In a parallel study, \cite{lee2020} simulated atmospheric circulation specifically for the brown dwarf orbiting white dwarf WD0137-349 with orbital period of about 2 hours using a GCM coupled with a dual-band grey radiative transfer scheme.  Their results are qualitatively similar to the our most rapidly rotating case, showing a narrowing equatorial superrotating jet, nearly geostrophic flows at mid-high latitudes and large day-night temperature difference. Although using a different GCM and radiative forcing setup, the agreement between two studies is encouraging.  A direct implication of our study is to understand and interpolate the phase curve observations of brown dwarfs in extreme close-in orbits around white dwarfs. Because these atmospheres might be coupled to the intrinsic magnetic field, it is interesting to constrain the role of possible magnetic drag by using   multi-wavelength phase curves and long-time monitoring to measure the dayside flux variability. An alternative way may be  the Doppler effect by winds on the high-resolution transmission spectroscopy, and this may  be able to constrain whether the flow is predominantly day-to-night or zonal \citep{showman2013b,louden2015,brogi2016,flowers2019}.  

{\ttt \cite{showman2015} also explored effects of increasing rotation (but non-synchronously rotating) in the circulation of hot Jupiters. Their fast rotating models exhibit weak equatorial superrotation or  no equatorial superrotation at all, while strong jets exist at high latitudes due to baroclinic instability. This is different to our models with short $\tau_{\rm rad, top}$, in which strong equatorial superrotation is prevalent no matter how fast they rotate. This is partly because the forcing mechanisms are different. In the fast rotating, non-synchronous models of \cite{showman2015}, the rapid westward migration of the irradiated pattern may disrupt the formation of a standing Matsuno-Gill pattern, therefore weakens the driving force for the equatorial superrotation. In addition, the equator-to-pole temperature gradient as a result of non-synchronous rotation generates baroclinic instability, and Rossby waves from high latitudes propagate to low latitudes and contribute to the westward acceleration on the zonal wind. These together imply a much weaker (or even no) equatorial superrotation and strong high-latitude jets in the fast rotation models of \cite{showman2015}. Whereas in our models, the steady, day-night forcing exists no matter how fast the model rotates. Therefore, there is always a standing Matsuno-Gill pattern that drives a strong equatorial superrotating jets no matter how fast the model rotates.  
}


The atmospheric temperature of BDs in tight orbits around WDs may be sufficiently high to thermally dissociate some fraction of {\tt the} molecular hydrogen, especially at low pressure where {\tt the gas absorbs the} UV flux from the WD. The dynamical effects of hydrogen dissociation and recombination, including chemical heat release/absorption and change of mean molecular weight  \citep{tan2019b}, would play an additional  role in shaping the atmospheric circulation.   The eddies associated with the instabilities could be more energetic, {\tt analogous} to {\tt the} effects of latent heat on baroclinic eddies {\tt in the} midlatitudes of Earth's atmosphere (e.g., \citealp{lapeyre2004}).  The size of eddies might be sufficiently large and the associated temperature perturbations {\tt sufficiently high for these effects to have observable consequences}.

Due to the inefficient day-night heat transport, {\tt the} temperature structure {\tt on the} nightside of BDts in tight {\tt orbits} around WDs may be {\tt similar} to that of field BDs, and optically thick silicate or iron clouds could form there. As has been demonstrated in \cite{tan2019}, the intrinsic variability driven by cloud radiative feedback may occur over the nightside of brown dwarfs and can induce high-amplitude temperature and cloud structure variation.  The heating/cooling rate associated with the variability could {\tt even} be comparable to that {\tt associated with} the day-night forcing.  Interactions between the global scale, day-night driven dynamics with the small-scale intrinsic variability driven by cloud radiative feedback are yet an exotic regime which deserves future exploration.

In this work the  circulation of  tidally locked atmospheres with increasing rotation rate has been investigated in the context of gas giants, i.e., {\tt fluid hydrogen planets with} no solid surface and {\tt of Jupiter's} radius.   {\tt Still, our study also has important implications for extremely close-in} terrestrial planets observed to date. {\ttt Understanding the climate dynamics of them will be complementary to those in or near the habitable zones around low-mass stars, which are synchronously rotating but with a rotation period of several to more than 10 days (e.g., \citealp{merlis2010,yang2014,carone2015, haqq2018}).  }   Several ultra-short-period terrestrial planets have been reported: KOI 1843.03 (4.245 hours of orbital period and hence the rotation period if it is tidally locked, \citealp{rappaport2013}), K2-137 b (4.3 hours, \citealp{smith2018}), Kepler-78 (8.5 hours, \citealp{winn2018}), LHS 3844 b (11 hours, \citealp{vanderspek2018, kreidberg2019}), {\ttt 55 Cancri e (17.7 hours, \citealp{demory2016}) and Kepler 10b (20 hours, \citealp{rouan2011})}.  Their primary atmospheres are likely blown away by stellar irradiation or stellar wind due to their extreme proximity, but the vigorous outgassing due to  high surface temperature may  help them to retain secondary atmospheres.  Although their measured {\tt radii} are much less than {\tt Jupiter's} radius, the equatorial deformation radius may still be a small fraction of their planetary radius depending on the thickness {\tt and} stratification of their atmospheres. {\ttt Effects of much lower surface gravity and presence of a solid surface likely do not affect the qualitative dynamical regime transition discussed in this study. As long as the equatorial Rossby deformation radius is much smaller than the planetary radius and $Ro\ll 1 $ at mid latitudes, the quasi-geostrophic regime naturally emerges.} Thus, based on {\tt our results}, we expect the thermal phase curves may be aligned with the secondary eclipse and the day-night temperature differences should be large for these extreme close-in terrestrial worlds. If the major atmospheric composition is rock or iron vapor, the {\tt atmospheres} may be able to condense {\tt on} the much cooler night side, and dynamical properties of non-dilute  condensable atmospheres \citep{ding2016, pierrehumbert2016,hammond2017} would pose further questions about the nature of the atmospheric circulation of this kind. {\ttt Note that for lava planets, rotation and tidal dissipation could affect the phase curve in addition to atmospheric flows (e.g., \citealp{selsis2013}). } 

{\ttt \cite{carone2015} pointed out that the climate regime transition from a single equatorial jet to more jets of tidally locked rocky planets occurs when the equatorial deformation radius is smaller than half of the planetary radius. \cite{haqq2018} suggested another interesting regime transition that could occur when the Rhines scale becomes smaller than the planetary radius. The smallest orbital (and hence the rotation) period of these studies are still comparable to a day. As mentioned above, it would be interesting  to examine the regime transition further to a rotation period down to several hours for tidally locked rocky planets. 
}

Brown dwarfs orbiting white dwarfs may host an intrinsic heat flux much larger than those of typical hot Jupiters, which could result in thermal perturbations {\tt caused by convection impinging against the base of the atmosphere} near the radiative-convective boundary \citep{showman&kaspi2013, zhang&showman2014, showman2019}. These perturbations are presumably small-scale, {\tt and they can drive} large-scale zonal jets and bands via {\tt wave-mean flow interactions} \citep{showman&kaspi2013, showman2019}. In this study we do not include such effects in order to keep a clean modeling framework. However, {\tt the} irradiated flux of these extreme close-in gas giants still dominates over intrinsic flux. For example, the irradiated equilibrium temperature of WD0137-349b \citep{casewell2015} is on the order of 2000 K, but the intrinsic flux may be on the order of only 1000 K in terms of effective temperature. Moreover, the efficiency driving atmospheric circulation by external day-night forcing is expected to be much larger than that driven by the internal flux. The former can be thought as a heat engine powered by day-night contrast with a typical estimated thermodynamic efficiency on the order of 10\% for hot-Jupiter-like atmospheres \citep{koll2018}. The circulation is more indirectly forced by intrinsic heat flux which has to first turn its {\tt convected interior heat flux} into atmospheric waves and turbulence, {\tt whence into} large-scale flow being forced by these eddies. Its efficiency is therefore significantly reduced by the two processes.   So we  expect that the circulation near the photosphere driven by day-night forcing would dominate over that by thermal perturbations in typical extreme close-in brown dwarfs around white dwarfs. At depth where the radiative timescale is long and  the day-night thermal forcing is weak,  zonal jets and banded structures can still emerge, although  the exact driving mechanism {\tt may differ from that in} isolated brown dwarf models, {\tt due to possible interactions between the interior convective flux and the external day-night forcing}.

\section{Conclusions}
\label{ch.conclude}

To summarize, we presented idealized general circulation models of {\tt synchronously rotating, strongly irradiated gas-giant planets} spanning a wide range of rotation period from 80 hours down to 2.5 hours to explore properties of the circulation with increasing rotation rate. {\tt Our motivation is to understand the dynamics of brown dwarfs in extremely tight, close-in (1--4 hour period) orbits around white dwarfs.}  Our primary results are as follows:
\begin{itemize}
	\item Decreasing rotation period (increasing rotation rate) causes a narrowing in the {\tt meridional width of the superrotating equatorial jet and the associated eastward-shifted hot spots, as well as the confinement to low latitudes of the standing Rossby wave (Matsuno-Gill) pattern.  At fast rotation rates, regions poleward of a few degrees latitude enter a regime of small Rossby number and geostrophic balance, which exhibits very different dynamics than the canonical, slow-rotating hot Jupiter regime (where Rossby number is order unity).}
	
	\item {\tt We showed that, on global scales, the day-night temperature difference increases with decreasing rotation period. Likewise, on global scales, the eastward offset of the dayside hot spot decreases as the rotation period decreases.}

	\item Thermal phase curves of our fast-rotating models show {\tt a close} alignment of the peak flux to phase 0.5 where a secondary eclipse would occur if the system were eclipsing, {\tt unlike the situation with typical hot Jupiters.  This result helps  explain the small offset of flux peak from phase 0.5 that has been observed in rapidly rotating BD+WD systems \cite[e.g.,][]{casewell2015}. Generally, as compared to hot Jupiters at similar irradiation levels, we expect brown dwarfs orbiting white dwarfs on short-period orbits to exhibit phase curves with larger day-night flux contrasts but smaller offsets of flux peak from secondary eclipse.  Atmospheric drag (if any) exerts a significant effect on the phase curves, although more realistic models are needed for a proper representation of the Lorentz force our drag is intended to parameterize.} 
	
	\item On rapid rotators, there exist {\tt a wealth of small-scale eddies and associated time variability at both low and high latitudes,} which are likely due to barotropic and baroclinic instability. {\tt Both types of instability may be important in shaping the mean atmospheric state, including the structure of the zonal jets and temperature.}  Meandering equatorial features are prevalent {\tt on the superrotating jet of} all models, {\tt probably} due to barotropic instability.  
	
	\item The equatorial superrotating jet is prevalent in models at all rotation period. The meridional half width of the equatorial jet scales closely with the equatorial deformation radius, as predicted by \cite{showman2011}. {\tt In rapidly rotating models, away from the equator, strong day-night forcing causes the generation of numerous, alternating eastward and westward zonal jets. These are analogous to the zonal jets caused by convection or baroclinic instabilities in more familiar Jovian settings \citep[e.g.][]{vasavada2005}, but here these jets form in response to a very different type of forcing.}  These off-equatorial jets scale well with the Rhines scale.
	
	\item We also explore models with long radiative timescale. In contrast to those with short radiative timescale, these models show much smaller horizontal temperature difference.  {\tt They do exhibit modest equator-to-pole temperature contrasts, and winds that are predominantly zonally aligned, but they do not exhibit the small-scale alternating jet structures prevalent on rapid rotators at short radiative time constant.}

	\item Our results have major implications for understanding atmospheric circulation of brown {\tt dwarfs} around white {\tt dwarfs} in extremely tight orbits, fast-rotating hot {\tt Jupiters} and terrestrial planets in ultra-short orbits. 	\\	
	
\end{itemize}

\acknowledgements
X.Tan acknowledges support from  the European community through the ERC advanced grant EXOCONDENSE (PI: R.T. Pierrehumbert).  A.P.S. acknowledges support {from} NSF grant AST 1313444. This work was completed with facilities provided by Lunar and Planetary Laboratory at the University of Arizona and the department of Physics at the University of Oxford.

\if\bibinc n
\bibliography{draft}
\fi

\if\bibinc y

\fi

\end{document}